\documentclass[12pt]{article}
\usepackage{graphicx}
\usepackage{float}
\usepackage{amsmath,amsthm,amssymb}

\newtheorem{conj}{Conjecture}

\newtheorem{prop}{Proposition}
\newtheorem{defn}{Definition}
\newtheorem{cor}{Corollary}
\newtheorem{lem}{Lemma}

\def\wt{\widetilde}
\def\wh{\widehat}

\newcommand{\Up}{\Upsilon}
\newcommand{\Th}{\Sigma}
\newcommand{\Om}{\Omega}
\newcommand{\Ga}{\Gamma}
\newcommand{\ga}{\gamma}
\newcommand{\na}{\nabla}

\newcounter{mnotecount}[section]

\title{Vacuum initial data on $\mathbb{S}^3$ from Killing vectors}
\author{Robert Beig$^{1}$,  Piotr Bizo\'n$^{2}$ and Walter Simon$^{1}$}
 \date{${}^{1}$Gravitational Physics, Faculty of  Physics, University of Vienna, Austria\\
${}^{2}$ Institute of Physics, Jagiellonian University, Cracow, Poland}
\begin{document}
\sloppy
\maketitle

\begin{abstract}
We  construct compact initial data of constant mean curvature $\wt{K}$ for Einstein's 4d vacuum equations with $\wh{\Lambda} = \Lambda - (\wt{K}^2/3)$ positive, where $\Lambda$ is the cosmological constant, via the conformal method.
To construct a transverse, trace-free (TT) momentum tensor explicitly we first observe that, if the seed manifold has  two orthogonal Killing vectors,  their symmetrized tensor product is  a natural TT candidate.  Without the orthogonality requirement, but on locally conformally flat seed manifolds  there is a  generalized construction for the momentum which also involves  the derivatives of the Killing fields found in work by Beig and Krammer \cite{BK}. We consider in particular  the round three sphere and classify the TT tensors resulting from all  possible pairs of its six Killing vectors, focusing on the commuting case  where the seed data are  $\mathbb{U}(1) \times \mathbb{U}(1)$  -symmetric. As to solving the Lichnerowicz equation,
we discuss in particular  potential ``symmetry breaking'' by which we mean that solutions  have
  less symmetries than the equation itself; we compare with the case of the  ``round donut''  of topology
  $\mathbb{S}^2 \times \mathbb{S}$.
In the absence of symmetry breaking,  the  Lichnerowicz equation for  a $\mathbb{U}(1) \times \mathbb{U}(1)$ symmetric  momentum  on $\mathbb{S}^3$  reduces to an ODE.
We analyze  distinguished  families of  solutions and the resulting data via a combination of analytical and numerical techniques.
Finally  we investigate marginally trapped surfaces of toroidal topology in our data.

\end{abstract}

\newpage

\section{Introduction}
We construct certain solutions of the initial value constraints in the compact case for the 4d Einstein equations with  cosmological constant $\Lambda$. Our tool is the conformal method (cf \cite{PC,JI} for recent reviews).  Simplifying the
general procedure, we set out  from a  ``seed manifold'' defined as follows.

\begin{defn}
\label{seed}
A {\bf Seed Manifold}  $({\cal M}, g_{ij}, L_{ij})$ consists of a compact 3-dim. manifold
$({\cal M},g_{ij})$ with smooth metric in the positive Yamabe class \cite{LP} and of a smooth tensor  $L_{ij}$  on ${\cal M}$ which is  transverse and traceless (TT; meaning $g^{ij}L_{ij} = 0  = \nabla^i L_{ij}$ where $\nabla$ is the covariant derivative).
\end{defn}
We wish to turn this into an initial data set of the following form:

\begin{defn}
\label{ids}
As {\bf CMC Initial Data} $(\wt {\cal M}, \wt g_{ij}, \wt K_{ij})$ (i,j,=1,2,3) for vacuum with cosmological constant $\Lambda$ we take a compact 3-dim. Riemannian manifold $\widetilde {\cal M}$ with smooth metric $\widetilde g_{ij}$ and
smooth symmetric (0,2) - tensor field $\widetilde K_{ij}$ which has constant trace
 $\widetilde g^{ij} \widetilde K_{ij} = \wt K = \text{const}$ and satisfies the constraints
\begin{equation}
\label{con}
\widetilde R   = \widetilde K_{ij} \widetilde K^{ij} - \widetilde K^2  + 2\Lambda, \qquad
\widetilde \nabla_i \widetilde K^{ij} = 0.
\end{equation}
Here $\widetilde \nabla$ and $\widetilde R$ are the covariant derivative and
the scalar curvature of $\widetilde g_{ij}$.
\end{defn}

To do so we need a smooth positive solution $\phi$ of the Lichnerowicz equation

\begin{equation}
\label{phi}
  \left( \Delta - \frac{1}{8} R \right) \phi = - \frac{1}{4} \widehat {\Lambda} \phi^5 - \frac{V^2}{8 \phi^{7}}
\end{equation}
where $\Delta = \nabla^i \nabla_i$ and $R$ are the
Laplacian and the scalar curvature of $g_{ij}$, and we have defined $\wh \Lambda = \Lambda - (\wt K^2/3)$ and $V^2 = L^{ij}L_{ij}$.
For every such $\phi$ the ``physical'' quantities
\begin{equation}
\label{ct}
\widetilde g_{ij} = \phi^4 g_{ij}, \qquad \widetilde K_{ij} -  \frac{1}{3} \wt g_{ij} \widetilde K  = \phi^{-2} L_{ij}
\end{equation}
 indeed satisfy the constraints (\ref{con}). The quantities $\widetilde g_{ij}$ and $\widetilde K_{ij}$ become the
  induced metric and second fundamental form (with constant mean curvature $\wt K$)  of a spacelike slice in the spacetime resulting from evolution under the $\Lambda-$ vacuum Einstein equations.
 For zero momentum $V \equiv 0$ Eq.  (\ref{phi})  is equivalent to the Yamabe problem  \cite{LP}.
We restrict ourselves to constructing data for which $\wh \Lambda > 0$; cf  \cite{PC}  for the case $\wh \Lambda \leq 0$.\\

\noindent
{\bf  Remark on notation}.  In what follows we abbreviate ``CMC initial data''  by ``data'' and  the ``seed manifold'' by ``seed''. Moreover,
``solutions''  $\phi$ of (\ref{phi}) are always understood to be smooth and positive. \\

Needless to say, the non-linearities of (\ref{phi}) are unpleasant. On the other hand, it is precisely due to this structure that the conformal method is capable in principle of generating ``interesting'' data from trivial seeds. In order to commemorate the
centenary of the work of our compatriot Friedrich Kottler \cite{FK}  (The Cracow region was with Austro-Hungary at that time)  we recall here the generation of time-symmetric ($\wt K_{ij} \equiv 0$) data for the Kottler (``Schwarzschild-de Sitter'') solution from a trivial seed.
Namely, we consider the ``round unit donut''
$\mathbb{S}^2 \times \mathbb{S}^1$ with a unit    $\mathbb{S}^2$ and an $\mathbb{S}^1$- circumference $B$ i.e.
\begin{equation}
\label{don}
ds^2 =  \left( d\xi^2 +   d\theta^2 + \sin^2 \theta d \varphi^2  \right)
\quad\mbox{where}  \quad \xi \in [0, B]
\end{equation}
and solve (\ref{phi}) for $V\equiv 0$.  Since $R = 2 $  there is the trivial solution $\phi \equiv \Lambda^{-1/4}$, but in addition we have  $k$ solutions of (\ref{phi}) if $B \in (2\pi k, 2\pi (k+1)]$ , ($k \in \mathbb{N}_0$)  which break the $\mathbb{U}(1)$ symmetry of the seed metric and of the equation. The concept of symmetry breaking used here is the obvious one, but see Definition 7 below for the formal statement. Accordingly, there are $k$ non-trivial  physical data $(\wt{\cal M},  \wt g_{ij})$  which contain $j$ pairs ($ 1 \le j \le k$)  of maximal and minimal  surfaces  (cf  \cite{RS}).  The resulting  spacetime (which Kottler of course obtained like Schwarzschild, namely  via a spherically symmetric ansatz to  Einstein's equations)   then contains $j$ pairs of static ``black hole'' and ``cosmological'' horizons.

In this  paper we focus on $\mathbb{S}^3$, which is another well-known example for the Yamabe problem.
In this case there is even a 4-(continuous-) parameter family of non-trivial (i.e. symmetry breaking) solutions of (\ref{phi}) with
$V \equiv 0$.
Curiously, however, none of these solutions leads to new geometry, i.e.  conformal rescaling just rescales the
round sphere in a non-trivial way. We recall this  in Sect. 4.2.

The key issue in the present work  is a natural way of constructing TT tensors on  seed manifolds with continuous isometries
(Killing vectors, KVs).
 The simplest case  consists of  any seed which enjoys  a pair of  orthogonal KVs\,  $\Pi$ and  $\Up$, since  their symmetrized tensor product
 \begin{equation}
 \label{TTT}
 L_{ij}^{\dashv} = \Pi_{(i} \Up_{j)}
 \end{equation}
 is TT.  Such orthogonal   KVs  are e.g. $\Pi = \partial/ \partial \xi$ and  $\Up = \partial/\partial \varphi$ on the donut (\ref{don}).
 For a pair of general  (but possibly parallel) KVs there is a
 generalisation  of (\ref{TTT}), which still yields a TT tensor  provided  $({\cal M}, g_{ij})$ is of constant curvature, i.e.
 $3 \, R_{ij} = R \,  g_{ij}$.  On the unit sphere (where $R = 6$)  in particular, it reads

\begin{equation}
\label{BKT}
L_{ij}^{(\Pi, \Up)} = \Pi_{(i} \Upsilon_{j)}  + (\mathrm{curl}\, \Pi)_{(i} (\mathrm{curl}\, \Up)_{j)}   -  \frac{1}{3} g_{ij}
\left(  \Pi^k \Upsilon_k +    (\mathrm{curl}\, \Pi)^k (\mathrm{curl}\, \Up)_{k} \right).
\end{equation}
\noindent
In this paper we define  ``curl''  via
\begin{equation}
\label{rotation}
(\mathrm{curl} \,\Pi)_i := \frac{1}{2}\,\epsilon_i{}^{jk} \nabla_j \Pi_k.
\end{equation}
which   is $1/2$ of the standard definition but  saves factors of 2 elsewhere.
We call (\ref{BKT})  ``Beig-Krammer-tensor'' in view of a more general construction \cite{BK} requiring just
 a conformal KV and a divergence-free vector, on any locally conformally flat manifold.

Restricting  ourselves now  to the round three sphere as a seed, our aim is to classify the momenta (\ref{BKT}) which arise from all possible pairs of KVs.
We  first note that for two generic KVs on $\mathbb{S}^3$  the momentum term (\ref{BKT})  (and hence the evolving spacetimes) will not have any symmetries whatsoever.
 As to classifying the special cases,  the key is the unique decomposition of any KV $\Up$ on  $\mathbb{S}^3$ (cf. Lemma 1)  in terms of its self-dual (sd)
 and antiself-dual (asd)  parts
\begin{equation}
\Up = \Om + \Th
\end{equation}
defined  via  $\mathrm{curl}\, \Om  =  \Om$ and  $\mathrm{curl}\, \Th  = - \Th$.
Our result reads as follows.
(We omit obvious statements  which result  from applying the  (anti-)symmetry between sd and asd items): \\

\noindent
{\bf Theorem 1.  (Simplified version; for full statement cf. Sect.~3}.) 

\noindent {\it Let $\Pi$ and $\Up$ be two  Killing vectors on $\mathbb{S}^3$ (possibly parallel). Then the following holds:
\begin{description}
\item[I. The $\Lambda-$Taub-NUT case. ]
If  $\Pi$ is self-dual,  and  if the anti-self dual parts of $\Pi$  and $\Up $ are proportional,  $L_{ij}^{(\Pi,\,\Up)}$ is   $\mathbb{SO}(3) \times \mathbb{U}(1)$ invariant.

\item[II. The Homogeneous case.]
If  $\Pi$ is self-dual,   $L_{ij}^{(\Pi,\,\Up)}$ is  $\mathbb{SO}(3)$ invariant.

\item[III. The $\mathbf{U}(1) \times \mathbf{U}(1)$    case.]
If  $\Pi$ and  $\Up$ commute,  $L_{ij}^{(\Pi,\,\Up)}$  is  $\mathbb{U}(1) \times \mathbb{U}(1)$ invariant.

\item[IV. The $\mathbf{U}(1) $ case.]  If  the antiself-dual parts of $\Pi$ and $\Up $ are proportional,
 $L_{ij}^{(\Pi,\,\Om)}$ is $\mathbb{U}(1)$ invariant.

\end{description}}

As to solving the Lichnerowicz equation (\ref{phi}) with momentum term,  the key result
 is due to  Premoselli \cite{BP}.
In essence (we recall the full statement in Sect.\,4.1)   it reads as follows, in terms of a  constant $b \in \mathbb{R}^ +$  extracted
arbitrarily from $V = b \, \overline V$: There exists a $b_{\star} \in (0,\infty)$  such that   (\ref{phi}) has at least two solutions for
$0 < b <  b_{\star}$, precisely one solution for    $b = b_{\star}$,  and    no solution for  $b > b_{\star}$.
While this result applies to seeds  without any symmetry restrictions, it settles in particular  existence and non-existence in the cases of present interest, namely $\mathbb{S}^3$ and $\mathbb{S}^2 \times \mathbb{S}$.

We are now particularly interested if the ``symmetry  breaking''  mechanism discussed above for the Yamabe problem ($b = 0$) persists for $b \neq 0$.  This is indeed the case in the above-mentioned  $\mathbb{S}^2 \times \mathbb{S}$-example  - in particular ``rotating Kottler data''    arise in a natural way by solving (\ref{phi}) on the donut (\ref{don}), with a momentum given by (\ref{TTT}) (cf.  \cite{BPS} and Sect.\,4.3 below). On the other hand,  on $\mathbb{S}^3$  it seems  that the symmetry breaking Yamabe solutions mentioned before,  and discussed at length in Sect.\,4.2,  do not survive the addition of any momentum term.
More precisely, our findings in the cases listed in Theorem 1 are as follows:
In the homogeneous case (which clearly includes $\Lambda$-Taub-NUT) we have $V^2 = \text{const}$.
From a Theorem by Brezis and  Li \cite{BL}, this implies $\phi = \text{const}$ for solutions of (\ref{phi}), which yields an algebraic  equation for $\phi$. On the other hand, for the   $\mathbb{U}(1) \times \mathbb{U}(1)$- invariant momenta of point III above,
the commuting Killing fields span a torus,  and $V$  depends only on one variable $x$ labelling a toroidal foliation  of $\mathbb{S}^3$. Combining  numerical techniques with analysis
 we claim  that there are no symmetry breaking solutions for $b\neq 0$.
  More precisely, we conjecture (cf. Conjecture~2 for which we give a partial proof)
 that $\phi$ only depends on $x$ as well, whence (2) reduces
to an ODE whose observed solutions are just a "Premoselli pair" for every $b < b_{\star}$ (cf. Conjecture 1).
We display numerical results in the cases that $\Pi$ and $\Up$  are orthogonal and parallel.

The  final  Section 5 deals  with marginally (outer)  trapped surfaces (MTSs, MOTSs) and marginally trapped regions (MTRs). 
The former  are two-surfaces ${\cal S}$ defined by   $\wt \Theta^{\pm} =  0$  for at least one of the  null expansions $\wt \Theta^{\pm}$ on ${\cal S}$, while the latter are regions bounded by MTSs.  Our motivation comes from
a recent  criterion for the ``visibility'' of such MOTSs and MTSs  from timelike infinity in asymptotically de Sitter spacetimes,
 Theorem 2.5. of \cite{CGL}. We first review this result and discuss it in detail  for  toroidal MOTSs and MTRs in  de Sitter spacetime.     We next note that  an extension of this discussion to  perturbations of de Sitter seems feasible 
 by virtue of Friedrich's general stability results  \cite{HF2};  we formulate a corresponding conjecture.   
 Finally we consider   the $\mathbb U(1) \times \mathbb{U}(1)$- symmetric data constructed in Sect.\,4.2 in the three special cases of $\Lambda-$ Taub-NUT, and for $\Pi$ and $\Up$  orthogonal and parallel. 
 These data form one-parameter families  in which we can locate  toroidal MOTSs analytically ($\Lambda-$ Taub NUT)
 or  numerically in the other cases.   While the orthogonal case only yields the same (Clifford-)  torus as de Sitter itself,
 the MOTSs are quite non-trivial in the parallel case. We leave a discussion of the "visibility results'' for such MOTSs and MTRs 
 in the near-de Sitter setting to future work. 

\section{The three sphere and its symmetries}

The subsequent discussion  of  $\mathbb{S}^3$ is adapted to our construction (\ref{BKT})
of TT momenta discussed in detail the next section, and  focuses accordingly on pairs of  KVs.   While our presentation is  largely coordinate independent (in this section, we require coordinates in the proof of Lemma 3 only)  coordinate expressions  are included occasionally to  increase clarity.  The most useful ones for our purposes  are the following.

We restrict ourselves to the unit sphere embedded in flat  $\mathbb{R}^4$,  viz.
\begin{equation}
\label{cart}
ds^2 = dz_1^2 + dz_2^3 + dz_3^2 + dz_4^2   \qquad z_1^2 + z_2^3 + z_3^2 + z_4^2 = 1
\end{equation}
 We define ``toroidal coordinates'' ($\tau$, $\gamma$, $\xi$)~~~ $ \tau \in (0, \pi/2), ~~~ \gamma, \xi \in (0, 2\pi)$  ~by
\begin{equation}
\label{tor}
 z_1  =  \sin \tau \sin \gamma,   ~~    z_ 2 = \sin \tau \cos \gamma, ~~
  z_3 = \cos \tau \sin \xi,   ~~   z_ 4 = \cos \tau \cos \xi  \qquad
   \end{equation}
  which yields
  \begin{equation}
  \label{tau}
  ds^2 = d\tau^2 + \sin^2 \tau  d \gamma^2 + \cos^2 \tau  d\xi^2.
  \end{equation}
The name originates in the toroidal foliation $\tau = \text{const}$.

 The Riemann and Ricci tensors and the scalar curvature are given by
\begin{equation}
R_{ijkl} = 2 g_{k[i} g_{j]l}\,,\hspace{1.5cm}R_{ij} = 2 g_{ij}\,,\hspace{1.5cm}R = 6.
\end{equation}
We write scalar products  as $\langle \Pi, \Up\rangle =  \Pi_i \Up^i$, the vector product as
 $(\Pi \times \Up)_i = \epsilon_i{}^{jk} \Pi_j \Up_k$ and  the commutator (Lie bracket)
$[ \Pi, \Up]^j  = \Pi^i \nabla_i \Up^j -     \Up^i \nabla_i \Pi^j   $ is also standard.
There are six independent KVs which satisfy
\begin{equation}
\na_i \na_j \Pi_k = - 2 g_{i[j}\Pi_{k]}.
\end{equation}

Next observe that $\mathrm{curl}$ defined by (\ref{rotation}) maps KVs into KVs and
satisfies $\mathrm{curl}\,\mathrm{curl} = 1 $ on KVs.
This leads to the following key definition

\begin{defn}
We call Killing vectors  $\Om, \Th$ selfdual (sd) or antiself-dual (asd) when they obey $\mathrm{curl}\,\Om = \Om$
resp. $\mathrm{curl}\,\Th = - \,\Th$.
\end{defn}

The Lie bracket of KVs can be written as
\begin{equation}
\label{Sophus}
[\Pi,\Up] = (\mathrm{curl} \,\Up) \times \Pi - (\mathrm{curl}\, \Pi)  \times \Up
\end{equation}
Thus KVs which are curls of each other commute. Furthermore it follows that

\begin{equation}
\label{Lie}
[\mathrm{curl}\,\Pi, \Up]=[\Pi,\mathrm{curl}\,\Up]\,,
\end{equation}
so that sd and asd  KVs also commute.
We next note the  identity
\begin{equation}
\mathrm{curl}\,(\Pi \times \Up) = \frac{1}{2}(\Pi\, \mathrm{div} \Up - \Up \,\mathrm{div} \Pi  -
[\Pi, \Up])
\end{equation}
valid for arbitrary vector fields $\Pi, \Up$, which obviously reduces to
\begin{equation}
\label{curl}
\mathrm{curl}\,(\Pi \times \Up) =   - \frac{1}{2} [\Pi, \Up]
\end{equation}
for KVs. Still for KVs, (\ref{Sophus}),  (\ref{Lie}) and (\ref{curl}) now imply
\begin{equation}
\mathrm{curl}\,[\Pi,\Up] = [\Pi, \mathrm{curl} \,\Up] = [\mathrm{curl}\,\Pi,\Up]
\end{equation}

The Cartan-Killing metric is proportional to
\begin{equation}\label{Cartan}
{\cal G} (\Pi,\Up) = \frac{1}{2}\left (\langle \Pi , \Up \rangle +
\langle \mathrm{curl}\, \Pi , \mathrm{curl}\, \Up \rangle  \right)
\end{equation}
Given Killing vectors $\Pi,\Up
$, the expression (\ref{Cartan}) is constant on $\cal{M}$.
It is positive definite and $\mathrm{curl}$ is self-adjoint w.r. to ${\cal G}$, i.e.
\begin{equation}
{\cal G}(\mathrm{curl}\,\Pi,\Up) = {\cal G}(\Pi,\mathrm{curl}\,\Up)
\end{equation}
It has thus real eigenvalues, namely $\pm 1$ (which we knew already) and we obtain the following result.

\begin{lem}
The Lie algebra of Killing vectors in $\mathbb{S}^3$ decomposes into a direct sum of self-dual and antiself-dual Killing vectors  satisfying respectively $\mathrm{curl}\, \Pi = \pm\, \Pi$. In other words we can write every Killing vector $\Pi$  uniquely as

\begin{equation}
\label{dec}
\Pi = \Om + \Th
\end{equation}

where $\Om$ is self-dual and $\Th$ is antiself-dual.
Moreover, we have
\begin{equation}
\label{norm}
 \| \Om \|^2 =  {\cal G}(\Om,\Om) =  \langle \Om, \Om \rangle = \text{const} \qquad
 \| \Th \|^2 =  {\cal G}(\Th, \Th) =  \langle \Th, \Th \rangle = \text{const} \end{equation}
\end{lem}

We note that the decomposition (\ref{dec}) is orthogonal w.r. to the Cartan-Killing metric,
  while there is no orthogonality w.r.t.  to $g_{ij}$:  $\langle \Om, \Th \rangle \neq 0$.

We  continue with another  straightforward result.

\begin{lem}

 Killing vectors $\Pi$ and $\Up$ on $\mathbb{S}^3$ commute iff the self-dual as well as the antiself-dual parts   (\ref{dec}) of $\Pi$ and $\Up$ are linearly dependent.
In terms of the decomposition (\ref{dec})  this means that (possibly after interchanging $\Pi$ and $\Up$)  there exist constants $u \in \mathbb{R}$ , $v \in \mathbb{R}$ with
\begin{equation}
\label{lindep}
 \Pi =  u \, \Om + \Th,   \qquad
 \Up = \Om + v \, \Th.
\end{equation}
\end{lem}

 \noindent
 {\bf Remark.} The statement above  includes in particular  all pairs involving  sd and asd   KVs
 (one has to put either  $u = v = 0$, or  $\Sigma = 0$, or  $\Omega = 0$).\\

\noindent
{\bf Proof.} Inserting the sd-asd decomposition
\begin{equation}
 \Pi =  \Om^{\Pi} + \Th^{\Pi}    \qquad    \Up =   \Om^{\Up} +  \Th^{\Up}
\end{equation}
into (\ref{Sophus}) and using $[\Pi,\Up]=0$, yields
\begin{equation}
\label{cross}
 \Om^{\Pi} \times \Om^{\Up}  = \Th^{\Pi} \times   \Th^{\Up}.
\end{equation}
But  (\ref{Sophus}) and (\ref{curl}) imply that the vector product of KVs preserves the sd and asd
 subspaces. This implies that both sides of (\ref{cross}) vanish which yields the result.
\hfill $\Box$ \\

We next choose bases  $\Om^A$ and $\Th^A$
(capital latin indices take values $1,2,3$; upper and lower indices mean the same)
which are orthonormal w.r. to the Killing-Cartan metric.
For the standard scalar product this implies
\begin{eqnarray}
\label{orth}
  & &  \langle \Om^A , \Om^B \rangle  =  \delta^{AB}, \qquad
     \langle \Th^A , \Th^B \rangle = \delta^{AB} \\
      \label{comm}
   &  & [\Om^A, \Om^B]  = -2 \epsilon^{AB}_{~~~C}\, \Om^C,  \qquad       [\Th^A, \Th^B]  = 2 \epsilon^{AB}_{~~~C}\, \Th^C, \qquad   [\Om^A, \Th^B]  = 0,
   \end{eqnarray}
where $\epsilon_{ABC}= \epsilon_{[ABC]}$, $\epsilon_{123} = 1$
and, of course, $\mathrm{curl}\,\Om^A = \Om^A $ and  $\mathrm{curl}\,\Th^A  = - \Th^A $.
Eq. (\ref{comm})  means that the sd and asd subspaces form $\mathbb{SO}(3)$ Lie algebras,
while all pairs  with opposite duality commute. The different signs in the commutators are  the natural convention in view of  (\ref{Sophus}) which implies
$ [\Om^A, \Om^B]  = - \Om^A \times \Om^B$   but  $[\Th^A, \Th^B] =   \Th^A \times \Th^B$. Then
the remaining freedom in  $\{ \Om^ A,\Th^B \} $ are $\mathbb{SO}(3)$ - transformations
${\cal O}^A_{~B}$,\, $\widehat {\cal O}^A_{~B}$ of the form
\begin{equation}
\label{rot}
\Om^ {A\prime}  = {\cal O}^A_{~B} \, \Om^{B},    \qquad
   \Th^{A \prime} = \widehat {\cal O}^A_{~B}   \, \Th^{B},  \qquad
   {\cal O}^A_{~B} {\cal O}_{C}^{~B}  = \delta^{A}_{~C}, \qquad
    \widehat {\cal O}^A_{~B} \widehat {\cal O}_{C}^{~B}  = \delta^{A}_{~C}.
\end{equation}
The vectors $\{ \Omega^A, \Sigma^A \}$ act transitively on the $\mathbb{S}^3$.

In terms of the coordinates introduced in (\ref{tor}) and in terms of
 ``contravariant'' components $ \Om^A =   \Om^{A\,i}(\partial/\partial x^i)$,
 $\Th^A =   \Th^{A\,i}(\partial/\partial x^i)$, ordered as $(\tau, \ga, \xi)$,  they read

\begin{eqnarray}
\label{om}
& & \Om^1  =  \left( \begin{array}{c}  \cos(\ga + \xi)  \\ -\cot \tau \, \sin (\ga + \xi)  \\
   \tan \tau \,\sin (\ga + \xi) \end{array} \right) \qquad
\Om^2 =  \left( \begin{array}{c}  \sin(\ga + \xi)  \\ \cot \tau \,\cos (\ga + \xi)  \\
- \tan \tau \,\cos (\ga + \xi) \end{array} \right) \qquad
 \Om^3  =  \left( \begin{array}{c} 0 \\ 1 \\1 \end{array} \right) \qquad \\
\label{th}
& & \Th^1  =  \left( \begin{array}{c} - \cos(\ga - \xi)  \\ \cot \tau \, \sin (\ga - \xi)  \\
 \tan \tau \,\sin (\ga - \xi) \end{array} \right) \qquad
 \Th^2 =  \left( \begin{array}{c} - \sin(\ga - \xi)  \\ -  \cot \tau \,\cos (\ga - \xi)  \\
- \tan \tau \,\cos (\ga - \xi) \end{array} \right) \qquad
\Th^3  =   \left( \begin{array}{c} 0 \\ -1 \\1 \end{array} \right)
\end{eqnarray}

We now proceed with a more subtle result.

\begin{lem}
\label{lemorth}
 Suppose the Killing vectors $\Pi$ and $\Up$ on $\mathbb{S}^3$ are orthogonal.
 Then either
\begin{enumerate}
\item  both are either self-  or  antiself-dual, or
\item there are self and antiself-dual KVs $\Om$ and  $\Th$ with  $\| \Om \| = \|\Th \|$ and a constant
$c \in \mathbb{R} \setminus \{0\}$
such that
\begin{equation}
 \Pi = \Om + \Th ,   \qquad  \Up = c(\Om - \Th).
\end{equation}

\end{enumerate}
\end{lem}

\noindent
{\bf Proof}: In terms of the decomposition
\begin{equation}
\Pi = c_A \Om^A + d_B \Th^B,  \qquad    \Up = \overline c_A \Om^A + \overline d_B \Th^B
\end{equation}
with constants $c_A, d_B$, $\overline c_A$, $\overline{d}_B$,
$h_{AB} =   (c_A \overline d_B +  \overline c_A d_B)$
the requirement reads
\begin{equation}
\label{sp}
  0    =  \langle \Pi ,\Th \rangle   = c^A  \overline c_A + d^A \overline{d}_A +
h_{AB}\langle \Om^A ,\, \Th^B \rangle.
\end{equation}
Using the explicit forms (\ref{om}), (\ref{th}) we obtain from (\ref{sp})
\begin{eqnarray}
&  0  &  =  c^A  \overline c_A + d^A \overline{d}_A   -  [h_{11} \cos(\ga + \xi)   \cos(\ga - \xi) + h_{21} \sin(\ga + \xi)  \cos(\ga - \xi)
+ \nonumber \\
& &  + h_{12} \sin(\ga - \xi)   \cos(\ga + \xi) +h_{22}\sin(\ga + \xi)   \sin(\ga - \xi)   ]  + \nonumber \\
& +   \sin (2 \tau) &  [ h_{13} \sin(\ga + \xi)  +  h_{31} \sin(\ga - \xi)   -  h_{32}   \cos(\ga - \xi)  - h_{23} \cos(\ga + \xi)]
 + \nonumber \\
& + \cos (2 \tau) & [-  h_{11}  \sin(\ga + \xi)  \sin(\ga - \xi)   +  h_{21} \sin(\ga - \xi)   \cos(\ga + \xi)  + \nonumber\\ & &
 +   h_{12} \sin(\ga + \xi)  \cos(\ga - \xi) -  h_{22}  \cos(\ga + \xi)   \cos(\ga - \xi)   +  h_{33} ]
 \end{eqnarray}
for all $(\tau,\, \ga,\, \xi)$.  From this we first conclude that each bracket vanishes. The next step shows that
 $h_{AB} \equiv 0$ which also implies  $c^A  \overline c_A + d^A \overline{d}_A = 0$.
Contracting now $h_{AB} \equiv 0$ with $c^A$ and $d^A$ we find the following:
Either all $c_A$ and  all $\overline c_A$, or all $d_B$ and all $\overline{d}_B$ vanish, which yields the first alternative of the Lemma.
On the other hand, in the generic case we have non-vanishing constants
 $e$, $f$ such that $\overline c_{A} = e. c_A$ and  $\overline d_A = f. d_A$, and
inserting this into $h_{AB} \equiv 0$ gives $ f = - e$.
Finally, inserting this into  $0 = c^A  \overline c_A + d^A \overline{d}_A =  e (c_A c^A - d_A d^A)$
yields $\|\Om\| = \| \Th\|$. This gives the stated result.  \hfill{$\Box$}  \\

We  note  an obvious corollary to the above Lemmas.
\begin{cor}
 Suppose the Killing vectors $\Pi$ and $\Up$ are orthogonal and commute.
Then only the second alternative of  Lemma $\ref{lemorth}$ applies.
 \end{cor}

\noindent
{\bf  Remark.}
The preceding discussion suggests the following definition. Let $\Om$ and $\Th$ be self- and antiself- dual KVs on $\mathbb{S}^3$, respectively.  We define  the ``toroidal pair'' of Killing vectors $\Ga$ and $\Xi$ via

\begin{equation}
\label{tp}
\Ga =   \left( \frac{\Om}{\|\Om\|} + \frac{\Th}{\|\Th\|} \right), \qquad
\Xi  = \left( \frac{\Om}{\|\Om\|} - \frac{\Th}{\|\Th\|} \right).
\end{equation}

The terminology  originates in  the fact that in toroidal coordinates (\ref{tor}), the tangents $\Ga^A = \partial/\partial \gamma$ and $\Xi^A = \partial/\partial \xi$) to the torus $\tau = \text{const}$ indeed form a toroidal pair.
 In general,  $\Ga$ and $\Xi $ are orthogonal, curls of each other, commute,  and each one is hypersurface orthogonal as  it satisfies $ \langle \Ga ,\mathrm{curl}\,\Ga \rangle = 0 =
\langle \Xi ,\mathrm{curl}\,\Xi \rangle$.
Furthermore,  $\Ga $ and $\Xi$ have zeros (``axes'')  aligned along
 mutually linked great circles of $\mathbb{S}^3$. In contrast,  the $\Om$'s and $\Th$'s  are neither hypersurface orthogonal nor do they have an axis, since they don't even have zeros.

Clearly, every  KV $\Pi$  enjoys a ``toroidal decomposition''  via
\begin{equation}
\label{tp1}
\Pi = \Om + \Th  =  p \left( \frac{\Om}{\|\Om\|} + \frac{\Th}{\|\Th\|} \right) +
q \left( \frac{\Om}{\|\Om\|} - \frac{\Th}{\|\Th\|} \right)  = p \, \Ga + q \, \Xi
\end{equation}
in terms of its self- and antiself- dual parts $\Om$ and $\Th$, and with
 numbers $p$ and $q$  given by
\begin{equation}
2 p = \|\Om\| +  \|\Th\|,      \qquad  2 q = \|\Om\| - \|\Th\|.
\end{equation}

 Note,  however,  that there is some asymmetry in the decomposition (\ref{tp1}) since the first term
  is always present (as $p > 0$) ,
 while the second term is absent if  $\|\Om\| = \|\Th\|$.
 Precisely such a special  toroidal pair occurs as  point 2 of Lemma 3.

\section{The Beig-Krammer tensor}

Throughout the section, $\Pi$, $\Up$ are KVs on $\mathbb{S}^3$,
with self- and antiself-dual parts denoted by
\begin{equation}
\label{dec1}
\Pi = \Om^{\Pi} + \Th^{\Pi}, \qquad
\Up = \Om^{\Up} + \Th^{\Up},
\end{equation}
and $ \{\Om^A, \, \Th^{B} \}$ are the bases in the respective subspaces as introduced
in (\ref{orth}), (\ref{comm}).

The task is now to  discuss  the symmetries of the Beig-Krammer-tensor defined in the Introduction (\ref{BKT}).
We formulate its key property as follows.

\begin{prop}
On any space of constant curvature, i.e. $3 R_{ij} = R g_{ij}$,  the following tensor is TT:
\begin{equation}
\label{BKT1}
L_{ij}^{(\Pi, \Up)} = \Pi_{(i} \Upsilon_{j)}  + (\mathrm{curl}\, \Pi)_{(i} (\mathrm{curl}\, \Up)_{j)}   -  \frac{1}{3} g_{ij}
\left(  \Pi^k \Upsilon_k +    (\mathrm{curl}\, \Pi)^k (\mathrm{curl}\, \Up)_{k} \right).
\end{equation}
\end{prop}
\noindent
{\bf Proof.} This is a special case of the Theorem in \cite{BK}; alternatively the result can be obtained by  direct calculation.
\hfill $\Box$
\\ \\
The following Lemma the proof of which is obvious from Lemma 1 is key for our discussion of this tensor.

\begin{lem}
In terms of the decomposition (\ref{dec1})
 the  tensor (\ref{BKT1}) reads
\begin{equation}
\label{BKT2}
L_{ij}^{(\Pi, \Up)} =L_{ij}^{(\Om)} +  L_{ij}^{(\Th)} =
2 \,  \Om^{\Pi}_{(i}\, \Om^{\Up}_{j)}  -
\frac{2}{3}  \langle   \Om^{\Pi} , \Om^{\Up} \rangle g_{ij}  +
 2 \,  \Th^{\Pi}_{(i}\, \, \Th^{\Up}_{j)}  -
\frac{2}{3}  \langle   \Th^{\Pi} , \Th^{\Up} \rangle g_{ij},
\end{equation}
which  reduces to
\begin{equation}
\label{BKT3}
L_{ij}^{(\Pi, \Up)} =  2 u \,  \Om_i \, \Om_j  -
\frac{2 u }{3} \|  \Om \|^2  g_{ij}  +  2 v \,  \Th_i \, \Th_j  -
\frac{2 v }{3} \|  \Th \|^2  g_{ij}
\end{equation}
in the commuting case (cf.  Lemma 2).
\end{lem}

The point of this Lemma is that the differential expression (\ref{BKT1}) in terms of $(\Pi, \Up)$
is replaced by the purely algebraic ones (\ref{BKT2}), (\ref{BKT3}). We note that in these expressions there is no mixing
between the sd and asd components.  This leads to  our key  classification  relating  properties of the KVs $\Pi$ and $\Up$
to the symmetries of $L_{ij}^{(\Pi, \Up)}$.\\

\noindent
{\bf Theorem 1.} ~{\it Let $\Pi$ and $\Up$ be two  Killing vectors on $\mathbb{S}^3$ (possibly parallel).
Then in terms of the decomposition   (\ref{dec1}) and the basis (\ref{orth}), (\ref{comm})  we find

\begin{description}

\item[I.  The $\Lambda-$Taub-NUT case.]
If  $\Up$ is self-dual, (i.e. $\Th^{\Up} = 0$),  and if $\Om^{\Pi} = u\,  \Om^{\Up}$ for some
constant $ u \in \mathbb{R}$, then   $L_{ij}^{(\Pi,\,\Up)}$ is invariant under the $\mathbb{SO}(3) \times \mathbb{U}(1)$
 action generated by $\{\Th^A, \, \Om^{\Pi}  \}$.

\item[II. The Homogeneous case.]

If  $\Up$ is self-dual, (i.e.  $\Th^{\Up} = 0$),  then  $L_{ij}^{(\Pi,\,\Up)}$ is invariant under the $\mathbb{SO}(3)$
 action generated by $\{\Th^A \}$.

\item[III. The $\mathbb{U}(1) \times \mathbb{U}(1)$ case.]
 If  $\Pi$ and  $\Up$ commute,   $L_{ij}^{(\Pi,\,\Up)}$ is  $\mathbb{U}(1) \times \mathbb{U}(1)$ invariant.
From Lemma 2 and Lemma 4, the invariance is generated  by  \{$\Om, \Th$\} unless one of these latter vectors vanishes,
 in which case the invariance group enlarges to
 $(\mathbb{SO}(3) \times \mathbb{U}(1))  \supset (\mathbb{U}(1) \times \mathbb{U}(1))$
 and yields $\Lambda-$Taub-NUT  data (cf.  I).


\item[IV. The Unitary case.]
If   $\Om^{\Pi} = c\,  \Om^{\Up}$ for some constant $ c \in \mathbb{R}$, then
$L_{ij}^{(\Pi,\,\Om)}$ is invariant under the $\mathbb{U}(1)$ action generated by  $\Om^{\Pi}$.

\end{description}}

\noindent
{\bf Proof}: 
The main statements ({\it  If...,then...}) of cases  I,  II and IV  are immediate consequences of Lemma 1, Lemma 4, and the commutation relations (\ref{comm}). In addition, the proof  that case  I  indeed produces   $\Lambda-$ Taub-NUT data is postponed to the Appendix.  As to  case III,  it is  obvious that  $L_{ij}^{(\Pi,\,\Up)}$  in the original form (\ref{BKT1}) is $\mathbb{U}(1) \times \mathbb{U}(1)$ invariant  under the action of its  commuting generators $\Pi$ and $\Up$, and hence under any linear combination thereof,  {\it unless these vectors are parallel}.  In  this special case the full invariance  still holds and follows from Lemmas 2 and 4 as stated
in III above. 
\hfill{$\Box$}\\

\noindent
{\bf Remarks.}
\begin{enumerate}
\item We have omitted obvious counterparts to the above statements which result from applying the (anti-)symmetry between
sd and asd items.

\item If $\Pi$ and $\Upsilon$ are orthogonal, then the data are either homogeneous (case
II) or $\mathbb{U}(1) \times \mathbb{U}(1)$ symmetric (case III), which follows immediately from the two cases of Lemma 3.
\item Concerning the $\mathbb{U}(1) \times \mathbb{U}(1)$ symmetric data,  there are  the following interesting special cases (in the notation of Lemma 2):
\begin{description}
\item[$\Lambda$-Taub-NUT:] Applies if one of the following holds: $\Sigma \equiv 0$, $\Omega \equiv 0$,  $u = 0$, $v = 0$, or $u = v = 0$.
\item[  the ``parallel'' case:]  $u \,v = 1$ (but neither   $\Sigma \equiv 0$ nor $\Omega \equiv 0$).
\item[the ``orthogonal'' case:] $u \, v = - 1$ and $ u  \|\Om\| = \|\Th\|$ (cf.  point 2 of Lemma 3).
\end{description}

 On these  data  we will focus our  discussion of the Lichnerowicz equation in Sect.\,4 below,
 and determine the marginally trapped surfaces in the data in Sect.\,5.

\item From the previous remark it is clear that case I  is  a special case of any other one, while III is a special case of IV.

\item Some converse of the above theorem holds as well, i.e. invariances of  $L_{ij}^{(\Pi,\,\Up)}$ imply statements
on $\Pi$ and $\Up$. The proof is non-trivial in case III only; we refrain from giving details.

\item Clearly, the list in the above theorem is not exhaustive, i.e. there is a generic case (no continuous symmetries) as well.

\end{enumerate}

\section{The Lichnerowicz equation}

\subsection{Existence, stability and symmetry}
We recall here  key results  \cite{BP} on  solving the Lichnerowicz equation (\ref{phi})
and on proving properties of its  solutions.   Since both our seed manifold as well as the momentum term $V^2$ enjoy
symmetries, it will in particular be important to examine the conditions under which the solutions
inherit or break these symmetries. We recall  from  \cite{BPS}
some definitions and results on  this issue. We then compare  their application to the {$\mathbb{S}^3$ and
 $\mathbb{S}^2 \times \mathbb{S}^1$ cases \cite{BPS}, respectively. We recall from the Introduction that
 a ``solution''  is always understood to be smooth and  positive.

\begin{defn}{\bf The linearized Lichnerowicz operator $L_{\phi}$ and  eigenvalue $\lambda$:}
\begin{equation}
\label{lin}
L_{\phi} \zeta := \left(- \Delta + \frac{R}{8} - \frac{5 \wh \Lambda}{4}
\phi^4 + \frac{ 7 V^2}{8 \phi^{8}}\right) \zeta = \lambda \zeta.
\end{equation}
\end{defn}

\begin{defn}{\bf  Stability of solutions and initial data sets.}
\begin{enumerate}
{\it \item A solution $\phi$ of (\ref{phi}) is
 strictly stable/stable/marginally stable/unstable/strictly unstable iff the lowest eigenvalue
 $\lambda$  is positive/nonnegative/zero/nonpositive/negative
respectively.
\item An initial data set  $(\wt {\cal M}, \wt g_{ij}, \wt K_{ij})$ is
stable iff  the solution $\phi \equiv 1$ is stable, and analogously for the other stability properties.}
\end{enumerate}
\end{defn}

We remark that stability of a data set implies that in fact {\it every solution} (generating that data from an an arbitrary seed)
 is  stable (cf. Lemma 1 of \cite{BPS}).  Again this extends to all stability properties.

\begin{prop}{\bf Uniqueness of stable solutions for convex potentials, cf.  \cite{LD}.}
If stable solutions of  (\ref{phi}) exist, they are unique.
\end{prop}
\noindent
 To see this, note that the potential term in Eq. (\ref{phi}) is strictly convex in the sense of  Proposition 1.3.1. of \cite{LD}. Thus 
 the latter result just requires  adaptation from the autonomous case to the present non-autonomous one,
 and from Dirichlet boundary conditions to the present compact case. Both generalisations are trivial.\\

\noindent
{\bf The Yamabe theorem  (cf.  \cite{Rick, LP}).} {\it  Let $({\cal M}, g_{ij})$ be compact and of positive Yamabe type.
Then (\ref{phi})  with  $V \equiv 0$  has at least one  solution $\phi$.}\\

\noindent
{\bf Premoselli's  theorem \cite{BP}. }
{\it Let $({\cal M}, g_{ij})$ be compact and of positive Yamabe type.
Writing $V = b. \overline V $ for a positive constant  $b$  and a function $\overline V \not\equiv 0$,
the following holds:\\
There exists~ $b_{\star} \in (0, \infty)$ such that Eq. \eqref{phi} has, for

\begin{description}
\item[ $b < b_{\star}$\,:]  more than one solution precisely one of which, $\phi_s$,  is strictly stable;
\item [$b = b_{\star}$\,:] a unique marginally stable solution;
\item [ $b > b_{\star}$\,:] no solution.
\end{description}
Moreover, the unique stable solution $b \rightarrow  \phi_s(b) $ for $b \in (0,b_{\star}]$
satisfies
\begin{enumerate}
\item $\lim_{b \rightarrow 0} \phi_s(b)  = 0$;
\item the map $b \rightarrow  \phi_s(b)$ is continuous and increasing in the sense that $\phi_s(b) <  \phi_s(b')$
for $b < b'$, everywhere on ${\cal M}$;
\item
every $\phi_{s}(b)$   is  minimal in the sense that $\phi_{s}(b) < \phi $ everywhere, for any other solution $\phi$.
\end{enumerate} }

This formulation combines Theorem 1.1, Proposition 3.1 (positivity), Proposition 6.1  (stability and minimality)
and Lemma 7.1 (continuity) of \cite{BP}. Note that Theorem 1.1. applies to a more general setting in which uniqueness of stable solutions need not hold; in the present case it does follow from Proposition 2 above.

\noindent

We turn now  to the symmetry properties of solutions.

\begin{defn}
{\bf Symmetric Lichnerowicz equation.}
 We call Eq. (\ref{phi}) symmetric  iff  $({\cal M}, g_{ij})$ and  $V^2$ are invariant under some
(discrete or continuous) isometry.
\end{defn}

Clearly this definition is {\it a priori} less restrictive than the invariance of $({\cal M}, g_{ij})$ and $K_{ij}$
used in the remaining part  of this paper.

\begin{defn}
{\bf   Symmetry inheritance/breaking.}
 A solution $\phi$ of a symmetric Lichnerowicz equation (\ref{phi}) inherits  a continuous symmetry $\Pi$
 iff  the corresponding Lie derivative satisfies ${\cal L}_{\Pi} \phi \equiv 0$  while otherwise it breaks the symmetry.
An  analogous  definition applies to discrete symmetries.
\end{defn}

\noindent
\begin{prop} {\bf (Symmetry inheritance/breaking; cf Proposition 2 and Corollary 1 of \cite{BPS}).}
The stable solutions $\phi_s$ of (\ref{phi})  inherit continuous and discrete symmetries.
\end{prop}

\noindent

Premoselli's theorem also implies that the solutions form branches parametrized by $b$.
Of particular interest are results which characterize the behaviour of these branches near their end points
$b = 0$ and $b = b_{\star}$.  The minimal stable branch indeed enjoys such a ``universal''
behaviour on either end;  the precise results are as follows\\

\begin{lem} {\bf (modified part of Proposition 4 of \cite{BPS}).}
 There is an $\epsilon > 0$ such that for all $b \in (b_{\star} - \epsilon, b_{\star})$  there is precisely one stable and one unstable solution.
\end{lem}

\noindent
\begin{lem} {\bf (modified Proposition 3 of \cite{BPS}).}
{\it For the minimal, stable solutions $\lim_{b \rightarrow 0} b^{-1/4}\phi_s$ is finite.}
\end{lem}

On the other hand,  the number and the properties of the unstable branches largely depend on the seed and on $V^2$,
which is revealed in particular by the examples discussed shortly.
Nevertheless,  some general information can be obtained via the implicit function theorem,  bifurcation theory, and  general results on elliptic PDEs.  We recall in particular that a necessary condition for a bifurcation to occur at some $b$ is that the linearized operator $L$ defined as (\ref{lin}) has a zero eigenvalue.

In the next Sect.\,4.2  we discuss the $\mathbb{S}^3$ case which we  compare in Sect.\,4.3  with the round unit
donut $\mathbb{S}^2 \times \mathbb{S}$ (see(\ref{don})) elaborated  in \cite{BPS}.
Noting that $R = 6$ in the former and $R = 2$ in the latter case, it proves useful to remove $\wh \Lambda$ from  (\ref{phi})
via the rescaling
\begin{equation}
\label{sc}
 \psi = (\frac{2\wh \Lambda}{R})^{1/4}\,  \phi, \qquad    W = 2\frac{\wh \Lambda V}{R}
 \end{equation}
which yields
\begin{equation}
\label{lichsc}
\left( \Delta - \frac{R}{8} \right) \psi + \frac{R}{8} \psi^5 + \frac{W^2}{8 \psi^7}=0
\end{equation}
 \noindent
 for any $R = \text{const} > 0$. Note  that $\psi \equiv 1 $ now solves  (\ref{lichsc}) for $W \equiv 0$.
In terms of these variables, the linearization (\ref{lin}) reads
\begin{equation}
\label{linpsi}
L_{\psi} \rho := \left(- \Delta + \frac{R}{8} - \frac{5 R}{8}
\psi^4 + \frac{ 7 W^2}{8 \psi^{8}}\right) \rho= \lambda \rho.
\end{equation}

\subsection{$\mathbb{S}^3$}
In this case equation \eqref{lichsc} becomes
\begin{equation}
\label{lichs3}
 \Delta \psi - \frac{3}{4} \left(\psi - \psi^5\right)  + \frac{W^2}{8 \psi^7}=0.
\end{equation}
When we use the coordinate system (\ref{tau}) with the substitution   $x = \cos(2 \tau), ~~~ x \in (-1,1)$, we obtain
  \begin{equation}
  \label{torus}
 ds^2 = \frac{dx^2}{4 (1 - x^2)} + \frac{1 + x}{2} d\gamma^2 +   \frac{1 - x}{2} d\xi^2\,
\end{equation}
which we will use occasionally in what follows.

 We discuss in turn the Yamabe case $W \equiv 0$, the case that $W$ is constant,  and the generic
  $\mathbb{U}(1) \times \mathbb{U}(1)$-symmetric  one.

\begin{description}

\item[a) $\mathbf {W \equiv 0}$:]
This case is well-known, cf. e.g. \cite{LP} for a review.  We recall the Yamabe theorem and its proof
 which provides the most instructive example of symmetry-breaking and is required for the analysis of the generic case. It is based on the existence of nontrivial conformal isometries of the standard 3-sphere.\newline
 By means of preparation let us start with the following observations: there is a 4-parameter family of solutions of the equation
\begin{equation}\label{sigma}
\nabla_i \nabla_j \sigma + g_{ij}\sigma = 0.
\end{equation}
Namely, these can be taken to be constant linear combinations of the Euclidean coordinates $(z_1,z_2,z_3,z_4)$ (see (\ref{cart})), restricted to $\mathbb{S}^3$.
As a corollary they satisfy
\begin{equation}\label{coroll}
(\Delta + 3)\sigma = 0,
\end{equation}
whence are the $n=1$ spherical harmonics on $\mathbb{S}^3$. Next observe that by virtue of (\ref{sigma}) the vector fields $\sigma^i = \nabla^i \sigma$ are conformal Killing vectors. They form a 4-dimensional linear space, but not a Lie algebra.
Note that the quantity  $(\nabla \sigma )^2 + \sigma^2$  is constant; we find it convenient to rescale $\sigma$ such that

\begin{equation}\label{scale}
(\nabla \sigma )^2 + \sigma^2 = 1.
\end{equation}
Each of the functions $(z_1,z_2,z_3,z_4)$  has  this property. Note finally that each solution of (\ref{sigma},\ref{scale}) can be characterized as follows: pick a ('reference' or 'north pole') point $P$ on $\mathbb{S}^3$ and require $\sigma|_P = - 1$, whence $P$ is a critical point due to (\ref{scale}). The function $\sigma$ will then monotonically increase along the flow of $\nabla^i\sigma$ while being  constant on 2-spheres.
It goes to zero on the equatorial sphere and then to $+1$ on the point antipodal to $P$, which is also a critical point.

\begin{prop} {\bf  (The Yamabe theorem on $\mathbb{S}^3$).}
The solutions of (\ref{lichs3}) with $W \equiv 0$ form a  4-parameter family given by
\begin{equation}\label{a}
\psi_a  = \sqrt{\frac{2\, a}{(1 + \sigma)a^2 + 1 - \sigma}}
\end{equation}
where $a \in \mathbb{R}^+$.
\end{prop}

\noindent
{\bf Proof}:
Checking that $\psi_a$ solves (\ref{lichs3}) with $W=0$ is a straightforward exercise based on (\ref{sigma},\ref{coroll}).
As the reference point (north pole) can be chosen arbitrarily on  $\mathbb{S}^3$,
the full family of solutions is in fact 4-parametric. For uniqueness recall a theorem by Obata (see \cite{MO}, \cite{LP}), which states that all rescalings of the standard metric
having the same constant curvature come, apart from a constant rescaling, from conformal isometries of $(\mathbb{S}^3, g_{ij})$. We state without proof that the functions $\psi_a$, up to a constant rescaling by $a^{1/2}$, do come from the conformal flow
$\Psi_t$ generated by $\nabla^i \sigma$ with $a = e^t$.\hfill {$ \Box$}

Suppose we choose the north pole for $\sigma$ on the limiting great circle $x = -1$ on which the toroidal foliation given by $x = \mathrm{const}$ is based.
Then
\begin{equation}\label{aligned}
\sigma = \cos\tau \cos \xi\,,
\end{equation}

and we observe the following:  while the Yamabe equation (\ref{lichs3}) with $W \equiv 0$ is invariant under the six-parameter family   of isometries of $\mathbb{S}^3$, its solutions (\ref{a}) are of the form $\psi(r) = \psi(\tau,\, \xi)$ -  hence in particular the invariance under  the  KV $\Xi =  \partial/ \partial \xi$ is broken.

In view  of Proposition 2, this symmetry breaking signals an  instability under conformal rescalings.
In the present context, in particular for $R = 6$, $W = 0$ and $\psi = 1$,  Eq.  (\ref{linpsi}) becomes
\begin{equation}
\label{rho}
- \Delta \rho =  (3 + \lambda) \rho\,.
 \end{equation}
As is well known, the spectrum of $-\Delta$ on $\mathbb{S}^3$ is $n(n+2)$, where $n\in\mathbb{N}_0$.
  Thus, the lowest eigenvalue  is $\lambda= - 3$. In terms of the coordinates (\ref{torus}), the higher eigenmodes either depend on $x$ only, or they result from excitation of the $\gamma$  and  $\xi $- modes on a fixed torus $x= \text{const}$.
Explicitly, with the separation ansatz   $\rho(x, \ga, \xi) =  \exp({i k \ga + i m \xi})\,\chi(x)$,  where $i = \sqrt{-1}$ and
$k,m \in \mathbb{Z}$, equation  (\ref{rho}) takes the form

\begin{equation}
\label{sep}
\left(-4 \frac{d}{d x} (1-x^2)\frac{d}{d x}
+\frac{2 k^2}{1 - x} + \frac{2 m^2}{1 +x}\right) \chi  = (3+\lambda) \chi.
\end{equation}
In particular, the second eigenvalue $\lambda = 0$ has multiplicity four with the associated eigenfunctions
\begin{equation}\label{e_k}
  \chi_1=\sqrt{1-x} \cos{\xi},\quad \chi_2=\sqrt{1-x} \sin{\xi},\quad \chi_3=\sqrt{1+x} \cos{\gamma},\quad \chi_4=\sqrt{1+x} \sin{\gamma}.
\end{equation}
Of course, these eigenfunctions correspond to the four directions of the general 4-parameter
 solution of the Yamabe problem described above.

\item[b) $\mathbf {W = \text{const} \neq 0}$.]

In terms of the classification theorem for $L_{ij}^ {(\Pi, \Up)}$  of Sect.\,3, this  case arises precisely for the homogeneous case II in Theorem 1,
 as follows from (\ref{BKT2}) and $\langle \Om, \Th \rangle \neq 0$.   (Recall, however, that this homogeneous case
overlaps with the other  cases of the Theorem).

In this case, Eq. (\ref{lichs3}) has obviously  constant solutions determined by the positive roots of the polynomial
\begin{equation}
\label{const}
 \psi^{12} - \psi^8 + \frac{1}{6} W^2 = 0\,.
 \end{equation}
 These  solutions  come in pairs for all $0 < W^2 < W_{\text{max}}^2=8/9$, in accordance with Premoselli's theorem quoted in the previous subsection.
 There now arises the question of uniqueness of these solutions, particularly in view  of the symmetry breaking exposed
 above for the case $W \equiv 0$. However, it turns out that  this ambiguity disappears as soon as $W$ is turned on, due to the following result.
 
\noindent
 {\bf Theorem (Corollary 1 of Theorem 1 of Brezis and Li \cite{BL})}
 {\it On $\mathbb{S}^3$, the Lichnerowicz equation  (\ref{lichs3}) with $W = \text{const}$ has only constant solutions.}

\item[c) $\mathbf {W = W(x)}$.]

We finally turn to the case of $\mathbb{U}(1) \times \mathbb{U}(1)$  symmetric data.
From (\ref{BKT3}) we obtain, in terms of the notation of Lemma 2,
\begin{equation}
 \label{right}
W_{(\Pi ,\Up)}^2=  \frac{1}{9} \wh \Lambda^2 V^2_{(\Pi, \Up)}=
  \frac{8 \wh \Lambda^2}{27} \left[  \left( u^2 \|\Om\|^4  -  u\,v \,\|\Om\|^2 \,\|\Th\|^2 +
v^2  \, \|\Th\|^4 \right ) +
 3 \,u \, v \, \langle \Om, \, \Th \rangle^2\, \right] .
\end{equation}

We now choose  $\Om$ and $\Th$ as follows
\begin{equation}
\label{choice}
\Om = \sqrt{\frac{3}{\wh \Lambda}}\,\Om^3\quad  \mbox{and} \quad  \Th = u\, \sqrt{\frac{3}{\wh \Lambda}} \, \Th^3,
\end{equation}

 where $ \Om^3$, and $\Th^3$ are elements of a basis defined via (\ref{orth}), (\ref{comm}).
 This choice is compatible with (\ref{lindep}) and no loss of generality in view of the remaining scaling ambiguity and the rotation freedom (\ref{rot}).
  In terms of adapted coordinates
(\ref{om}), (\ref{th}) with $x = \cos(2 \tau)$  this entails $\langle \Om, \Th \rangle = 3\wh \Lambda^{-1}  u \, x $, and
simplifies   (\ref{right})  as follows
\begin{equation}
 \label{Wspec}
W_{(\Pi ,\Up)}^2=   \frac{8}{3} u^2  \left( 1 - u\, v  + u^2 \, v^2  \right) + 8 \, u^3\, v \, x^2.
\end{equation}

As to solving the  Lichnerowicz equation (\ref{lichs3}) we conjecture that, as for $W = \text{const} \neq 0 $ and in accordance with Premoselli's theorem quoted above,
 there is exactly one pair of solutions for $0<b<b_{\star}$. Below we split this conjecture into an ODE and a PDE part, and formulate it for arbitrary
$W^2(x) = b^2 \overline W^2(x)$ rather than for the special form (\ref{Wspec}). We call a solution even if $\psi(x) = \psi (-x)$.

\begin{conj}
 The  Lichnerowicz-ODE, which results from  (\ref{lichs3})  by assuming that $\psi = \psi(x)$, viz.
\begin{equation}
\label{lichx}
4 (1 - x^ 2)\,  \frac{d^2  \psi}{d x^2}  - 8 x\,  \frac{d \psi}{dx} - \frac{3}{4} (\psi - \psi^5) + \frac{b^2 \overline W^2(x)}{8 \psi^7} = 0
\end{equation}
has for every $0  < b \le b_{\star}$ a unique stable, even solution and a unique unstable,  even solution which coincide at $b = b_{\star}$.
For $b \rightarrow 0$ the solutions on the stable branch tend to zero like $b^{1/4}$, while the unstable ones converge to $\psi \equiv 1$.
\end{conj}

{\bf Partial proof and numerical evidence.} For the stable branch, the result follows from Premoselli's theorem and Proposition 2 and Lemma 6 above. For the unstable branch an adaption of Premoselli's theorem (in terms of suitably restricted function spaces) might still  apply. Alternatively, the implicit  function applied to the linearized ODE operator (the ODE restriction of (\ref{linpsi}))  would guarantee existence and uniqueness directly as long as this operator had a trivial kernel. This is easily verified for $b=0$ because the linearized Yamabe operator around $\psi=1$, given by $L=-\Delta -3$, has a trivial kernel when restricted to functions depending only on $x$, hence it holds for small $b$. It also  holds for $b$ near $b_{\star}$ by virtue of Lemma~5.
In the intermediate range we rely on numerical observations. \hfill $\Box$

 As  to the full Lichnerowicz equation \eqref{lichs3} with $b>0$, we now discuss non-existence of symmetry breaking bifurcations, first from the Yamabe solutions and then from the unstable ODE branch.

 \begin{prop}
 Equation \eqref{lichs3} has no symmetry breaking solutions bifurcating at $b=0$ from the zero eigenvalue of the four-parameter family of the Yamabe solutions.
 \end{prop}

{\bf Proof.}
We already know that for $W=W(x)$ and small $b$ there exists an ODE solution of equation \eqref{lichs3}
 of the form $\psi=\psi(x)=1+b^2 \psi_1(x)+\mathcal{O}(b^4)$.
To see if there are  other solutions bifurcating from $\psi=1$ at $b=0$, we seek them in the form
\begin{equation}\label{ansatz}
  \psi=1+b v + z,
\end{equation}
where $v\in N$ (kernel of $L$)  and $z\in N^{\perp}$ is of the order $\mathcal{O}(b^2)$. We recall that $N=\text{span}\{\chi_1,\chi_2,\chi_3,\chi_4\}$, where $\chi_k$ are given in \eqref{e_k}.
Substituting the expansion $z=b^2 z_2 + b^3 z_3+...$ into \eqref{lichs3}, at the order $\mathcal{O}(b^2)$ we get
\begin{equation}\label{b2}
  L z_2 = \frac{15}{2} v^2 +\frac{1}{8} \overline W^2.
\end{equation}
Since $v^2$  is orthogonal to $N$, by the Fredholm alternative the solution  $z_2$  exists iff
\begin{equation}\label{ort1}
(\overline W^2,\chi_k)=0,\quad k=0,...,4,
\end{equation}
where $(,)$ denotes the $L^2$-inner product on $\mathbb{S}^3$.
Assuming that the orthogonality conditions \eqref{ort1} hold,  at the order $\mathcal{O}(b^3)$ we get
\begin{equation}\label{b3}
  L z_3 = 15 v z_2 +\frac{15}{2} v^3 -\frac{7}{8} v \overline W^2.
\end{equation}
Let $v=\sum\limits_{k=1}^4 c_k \chi_k$. By the Fredholm alternative, the coefficients $c_k$ are constrained by the orthogonality conditions
\begin{equation}\label{ort}
  (15 v z_2 +\frac{15}{2} v^3 -\frac{7}{8} v \overline W^2, \chi_k)=0.
\end{equation}
In general, this will  give a system of four cubic polynomial equations for the coefficients $c_k$, however in our case  the cubic terms vanish identically and  we are left with the trivial linear equations $c_k=0$ (the vanishing of the cubic terms is  a consequence of  existence of the 4-parameter family of solutions for $b=0$; more precisely, for $W=0$ the $v$ and particular solutions $z_2,z_3,...$ correspond to the Taylor series expansion of this 4-parameter family). This excludes  symmetry-breaking solutions that bifurcate from $\psi=1$ at $b=0$, thereby proving the uniqueness of the $x$-dependent continuation (unstable branch) in $b$ of the trivial solution $\psi=1$.

Thanks to
 the conformal symmetry for $b=0$, an analogous argument proves the absence of bifurcations from the whole 4-parameter family of Yamabe solutions. \hfill $\Box$

 \begin{conj}
{\it Equation (\ref{lichs3}) has no symmetry breaking solutions bifurcating from the unstable ODE branch for  $b > 0$.}
\end{conj}

{\bf Partial proof and numerical evidence.}
We were not able exclude bifurcation and corresponding symmetry breaking by general theorems
such as  those of   Brezis and Li \cite{BL} used for $W = \text{const}$,  or by the results of
Jin, Li and Xu  \cite{ JLX} employed in \cite{CG}.
We rather have to resort to numerical evidence.
In particular we will  consider now  the eigenvalue problem for the linearized operator on the unstable branch
\begin{equation}\label{sep2}
\left(-4 \frac{d}{d x} (1-x^2)\frac{d}{d x}+\frac{3}{4} (1-5\psi^4)+\frac{7 W^2(x)}{8 \psi^8}
+\frac{2 k^2}{1 - x} + \frac{2 m^2}{1 +x}\right) \chi  = \lambda \chi
\end{equation}
 for special choices of $W(x)$.

 We  focus on  the  three cases listed in Remark 3 after Theorem 1,  namely $\Lambda$-Taub NUT (where we set $v = 0$), and the parallel ($u\,  v = 1$) and the
orthogonal ($u\,  v = - 1)$ cases.  Note that (\ref{choice}) is consistent with this Remark.

For the respective  momentum densities $W_{\Join}$, $W_{\parallel}$ and $W_{\bot}$
we obtain from (\ref{right})
\begin{equation}
\label{3Ws}
 W_{\Join}^2 =  \frac{8}{3}u^2,  \qquad  W_{\parallel}^{2} =  8 u ^2 (\frac{1}{3}  + x^2), \qquad
 W_{\bot}^{2 } =  8 u ^2 (1  -  x^2)\,.
\end{equation}
We can now consider  $b = u$  as  scaling parameter in  Premoselli's theorem, which then in particular  implies existence of solutions up to a maximal value $u_{\star}$.
Recall from the previous subsection b) that in the $\Lambda$-Taub NUT case where $W_{\Join} = \text{const}$  we have only constant solutions (see the upper diagrams in Fig. 1).
As to the other cases,  numerics and perturbative calculations show that along the unstable branches  the lowest eigenvalues grow monotonically from $-3$ at $u=0$ to $0$ at $u_*$
while all higher eigenvalues remain positive. This absence of zero modes supports the above Conjectures.
The  ODE branches are plotted  in Fig.~1.  \hfill $\Box$

We finally remark that an elementary perturbative calculation gives the following approximations for the {\it unstable solutions}
for small values of $u$
\begin{equation}\label{psi_pert}
\psi_{\parallel}(x)=1-\frac{5-x^2}{21}\,u^2+\mathcal{O}(u^4),\qquad
 \psi_{\bot}(x)=1-\frac{3 x^2+13}{63}\,u^2+\mathcal{O}(u^4).
\end{equation}
The corresponding eigenvalues (\ref{sep2})  can be obtained  perturbatively as well.
\vskip 0.2cm
\noindent {\bf Remark.} It is known from point 2. in Premoselli's theorem quoted above in section~4.1 that solutions on the stable branch are pointwise strictly monotonically {\it  increasing} with $u$. We observe numerically  that the solutions on the unstable branch
are pointwise strictly monotonically {\it decreasing} with $u$ (for small $u$ this follows from the perturbative solutions, cf. \eqref{psi_pert}). If proven, this would imply that the potential term in \eqref{sep2}
is strictly monotonically increasing with $u$ and consequently the same holds for the eigenvalues. This would prove that all the eigenvalues but the lowest one are positive, thereby excluding symmetry breaking bifurcations.

\begin{figure}[H]

\includegraphics[width=0.35\textwidth]{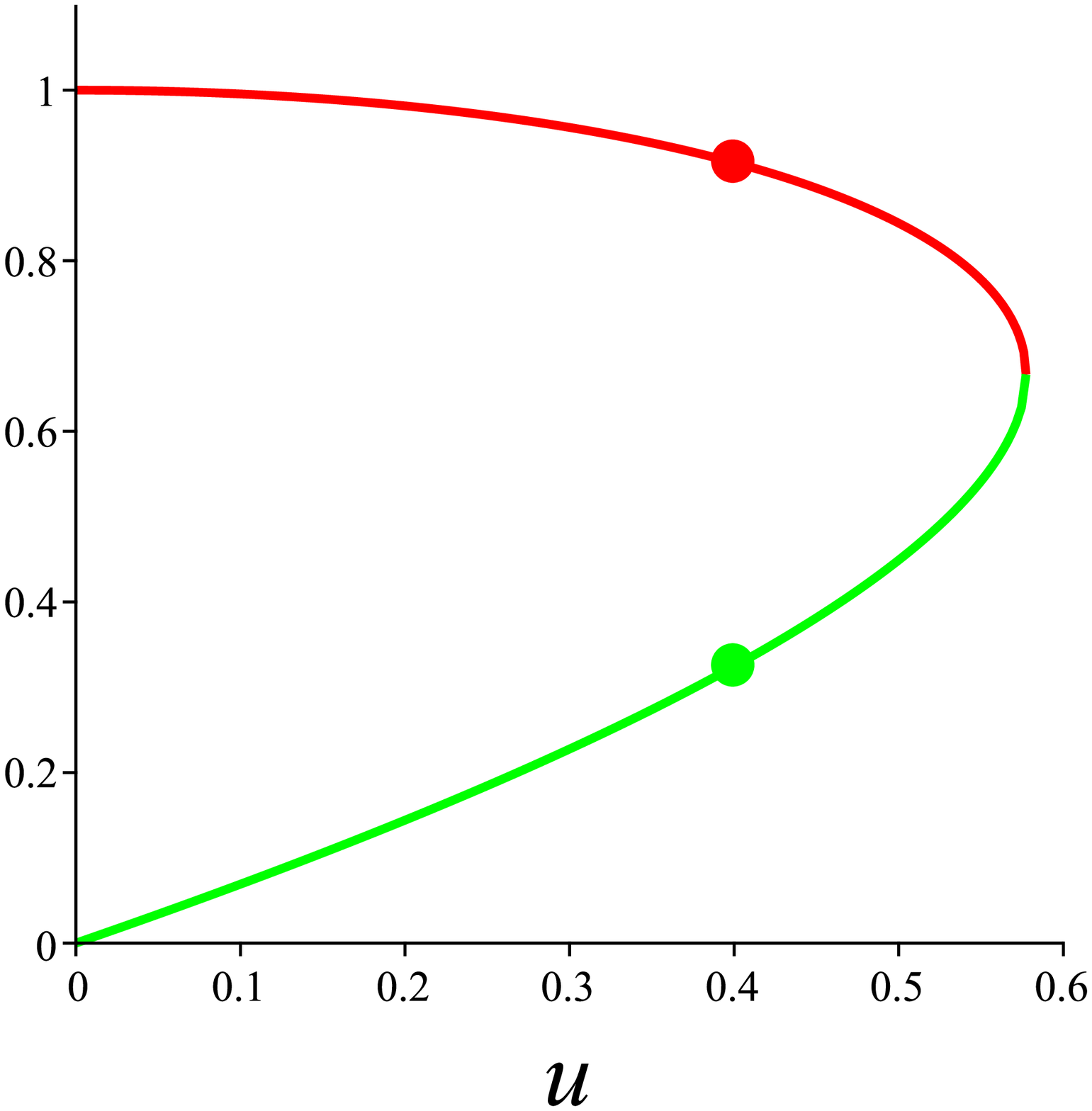}\hspace{2cm}
\includegraphics[width=0.35\textwidth]{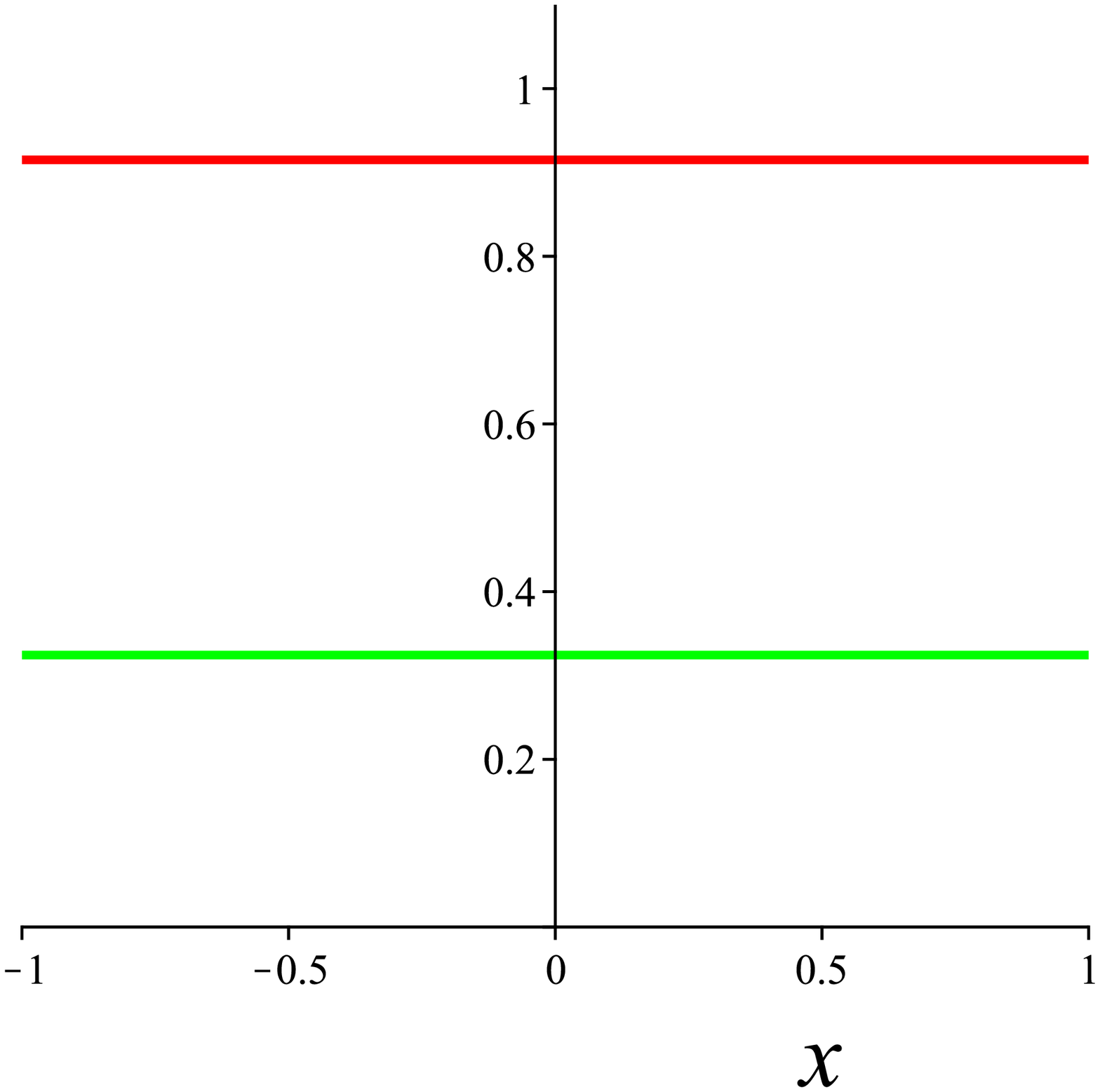}
\put(-430,78){\large $\psi_{\Join}^4$}
\put(-125,78){\large $\psi_{\Join}^4$}

\includegraphics[width=0.35\textwidth]{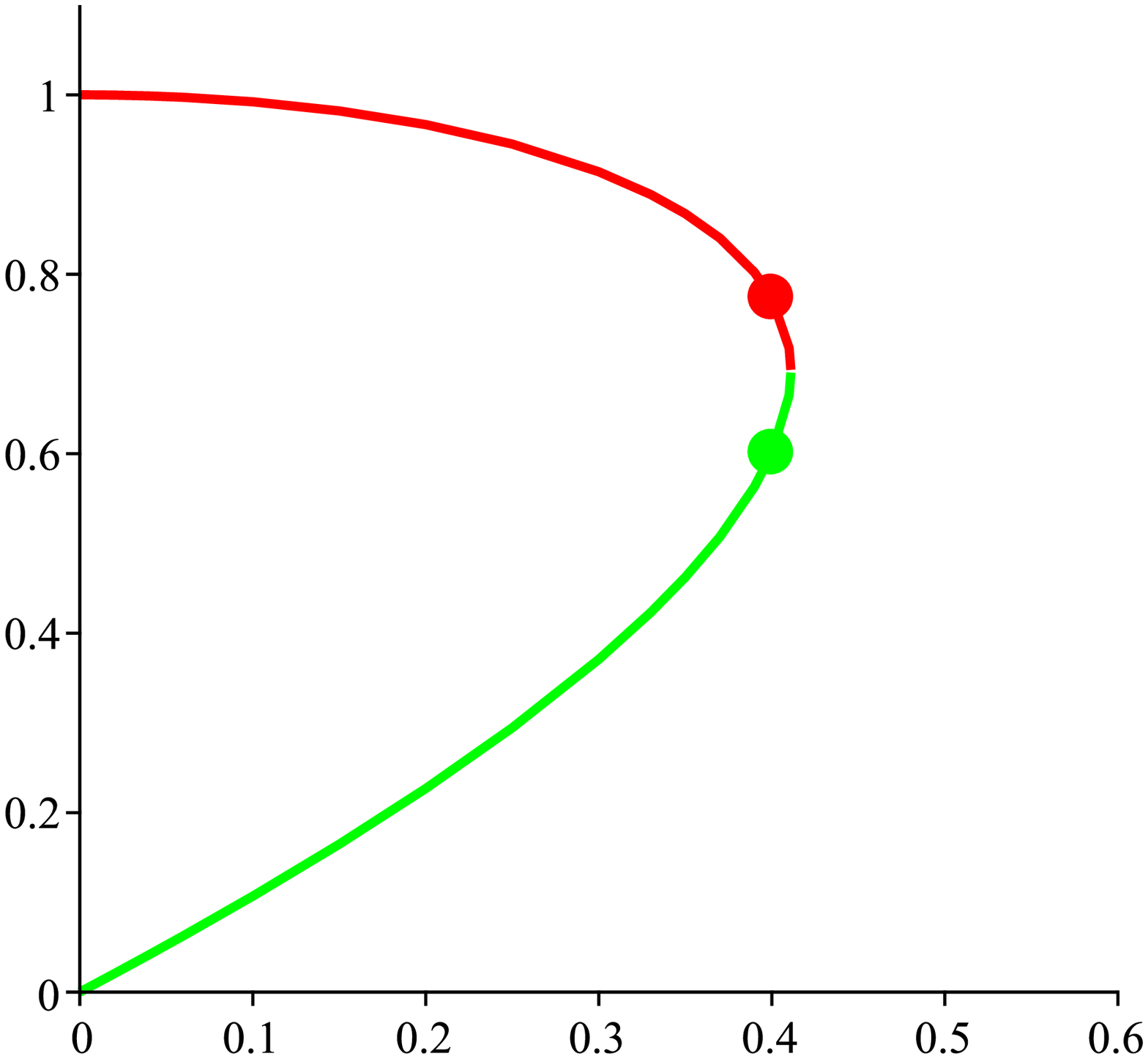}\hspace{2cm}
\includegraphics[width=0.35\textwidth]{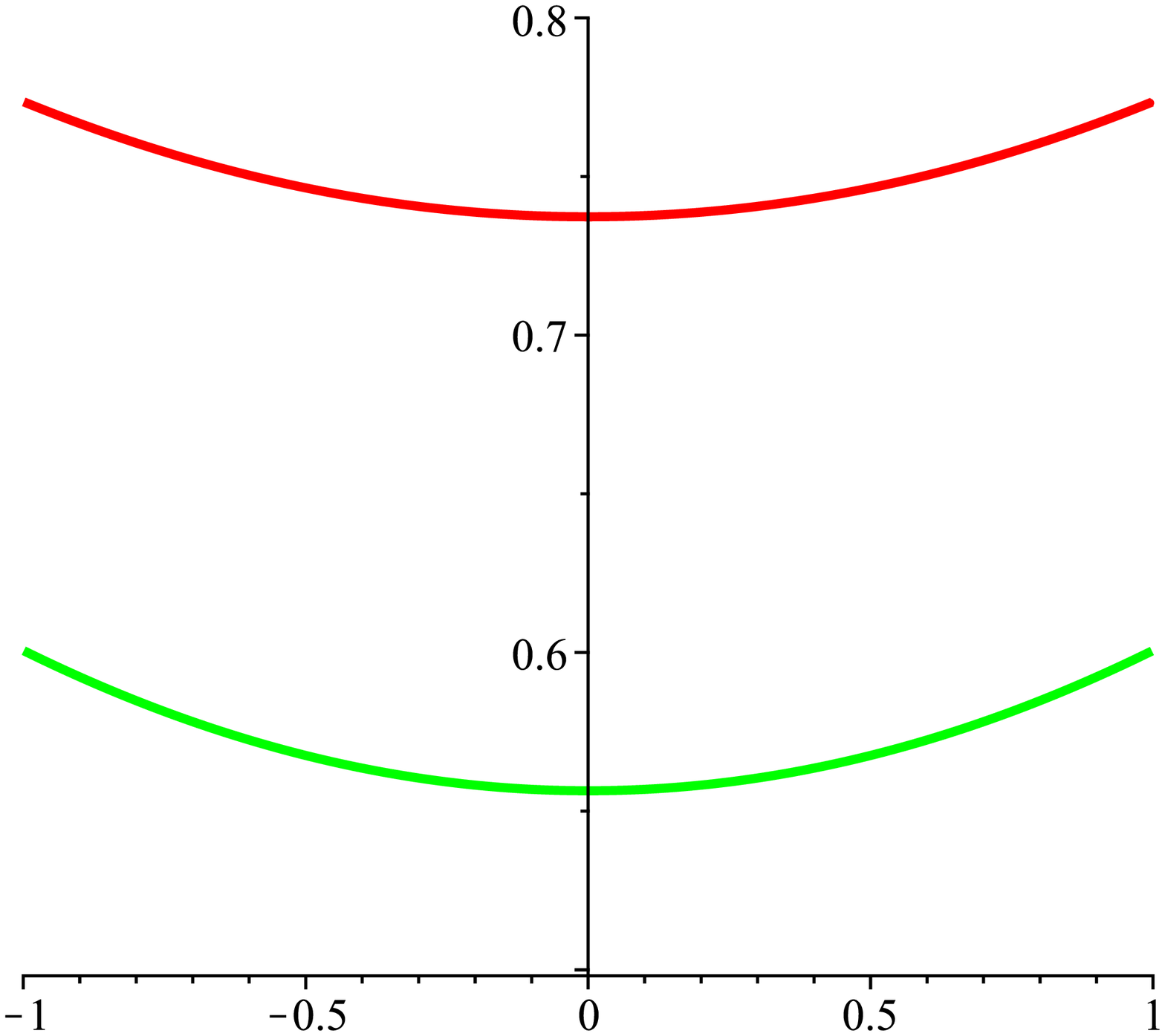}
\put(-450,78){\large $\psi_{\|}^4(\pm 1)$}
\put(-125,78){\large $\psi_{\|}^4$}

\includegraphics[width=0.35\textwidth]{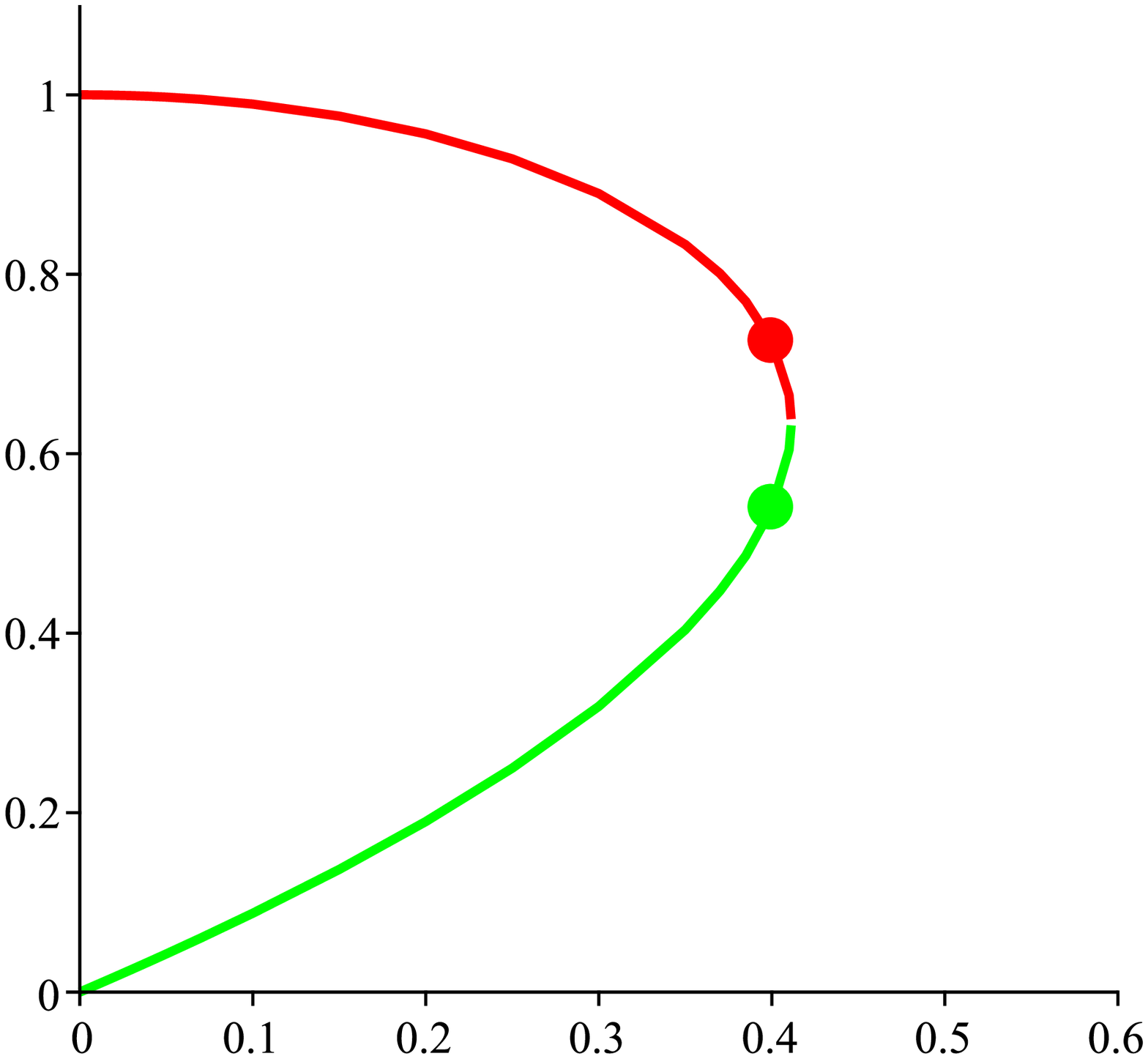}\hspace{2cm}
\includegraphics[width=0.35\textwidth]{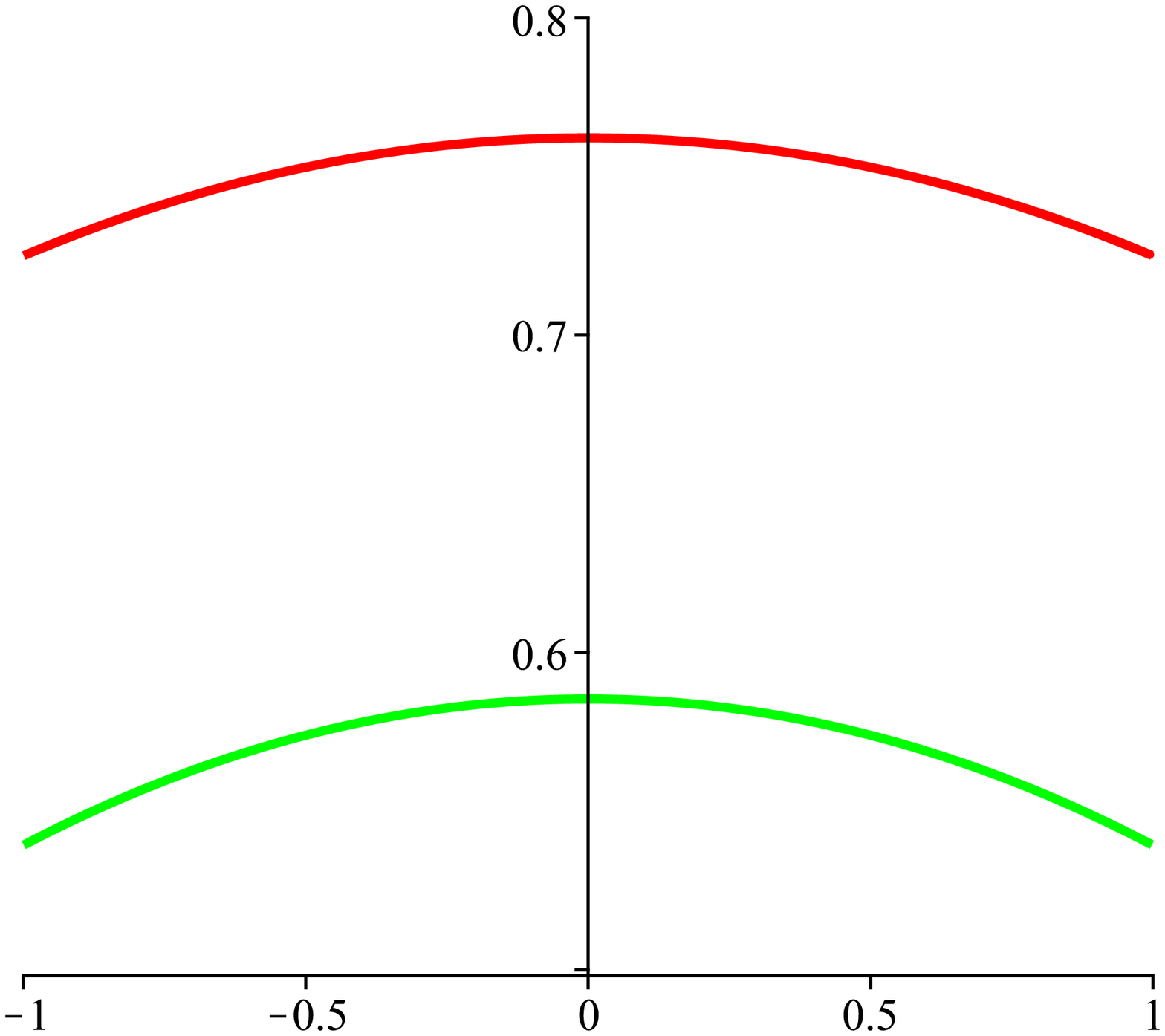}
\put(-450,78){\large $\psi_{\bot}^4(\pm 1)$}
\put(-125,78){\large $\psi_{\bot}^4$}
\vspace{1cm}
\caption{The diagrams on the left show the stable (green) and unstable (red) branches of solutions of the Lichnerowicz-ODE equation (\ref{lichx}), for the  choices $W = W_{\Join} = \text{const}$, $W(x) = W_{\parallel}(x)$ and $W(x) = W_{\bot}(x)$   given by (\ref{3Ws}); in the latter two cases,  there are plotted the values of $\psi^4$ at the polar circles $x = \pm 1$ against $u$. For a sample value  $u = 0.4$  (indicated by dots) the diagrams on the right show the respective functions $\psi^4(x)$. }
\end{figure}

\end{description}

\subsection{$\mathbb{S}^2 \times \mathbb{S}^1$}

It is instructive to compare the data constructed above with similar ones of topology
$\mathbb{S}^2 \times \mathbb{S}^1$.
In this case the  seed manifold is the donut (\ref{don}) whose symmetry group is obviously
    $\mathbb{SO}(2) \times \mathbb{U}(1)$. We consider here Eq. (\ref{lichsc}) for $R = 2$ and distinguish,
 in analogy with the $\mathbb{S}^3$ case,
    momenta  with $W \equiv 0$,
 $W = \text{const}$ and axially symmetric ones, i.e.  $W = W(\theta)$.

\begin{description}

\item[a) $\mathbf {W \equiv 0}$:]
 As we sketched in the introduction,   the Yamabe- case is already non-trivial for such data  since ``symmetry breaking
generates black holes'' - in particular, the family of time-symmetric Kottler data ($k$ solutions if $B \in (2\pi k, 2\pi (k+1)]$)  arises via breaking the $\mathbb{U}(1)$ symmetry  of the donut.  We refer to \cite{RS} for details.

\item[b) $\mathbf {W = \text{const} \neq 0}$.]
 In this case, the constant solutions are determined by the positive roots of the polynomial
\begin{equation}
 \psi^{12} - \psi^8 + \frac{1}{2} W^2 = 0,
 \end{equation}
which exist up to a maximum value  $W_{\text{max}}^2 = 8/27$. However, in contrast to the
$\mathbb{S}^3$ case,  there are also symmetry-breaking solutions;  we refer to Chru\'sciel \& Gicquaud \cite{CG}.
On the other hand, these authors show  that $\phi = \phi(\xi)$, i.e. the solutions are necessarily still $\mathbb{S}^2$-spherically symmetric.

\item[c) $\mathbf {W = W(\theta)}$.]
In \cite{BPS}  we considered a 3-parameter family of ``Bowen-York'' data which endows all $\mathbb{S}^2$ sections with an angular momentum of arbitrary magnitude and direction.  We recall that the 10-parameter family of Bowen-York-data \cite{BY} was originally defined on flat space,  but the definition carries over straightforwardly to the present locally conformally flat setting.
Analyzing then  the Lichnerowicz equation  via  Premoselli's theorem and numerically reveals a rich structure
of rotating data, which  exist up to a limiting angular momentum $J_{\star}$.  Among them are
both $\mathbb{U}(1)$-symmetry-preserving as well as  -breaking ones,  corresponding to their stability properties \cite{MK}.
Although the data with broken symmetry very likely contain ``black holes'' in the sense of marginally trapped surfaces
(as it is the case without rotation),  this is unproven.

We  show how the momenta  considered in \cite{BPS} are related to the scheme of  Sect. 3.  above.
To reinterpret this family of data in the present context, we  consider, in the coordinates (\ref{don}),
 the orthogonal, commuting  Killing vectors
\begin{equation}
  {\Xi}= \frac{\partial}{\partial \xi}, \qquad
{\Phi} =   \frac{\partial}{ \partial \varphi}.
\end{equation}
We note that, in  contrast to $\mathbb{S}^3$,  the present
$\mathbb{S}^2 \times \mathbb{S}$
 space is not of constant curvature,
whence the construction of  TT tensors described in Sect 3  does  not apply. Nevertheless,  we recall from Eq. (\ref{TTT})  of  the Introduction that the symmetrized  tensor product of $\Phi$ and $\Xi$, viz.
\begin{equation}
L_{ij} =  6 J \,\Xi_{(i}\, \Phi_{j)}  \quad  \mbox{where} \quad    J  = \frac{1}{8 \pi} \int_{\cal S} L_{ij} \Phi^i dS^ j
\end{equation}
is  a TT tensor on any background.  The (``Komar''-) angular momentum $J$
is  conformally invariant and the same for all  compact 2-surfaces ${\cal S}$ within a given homology class;
here in particular for any spherical surface.
As ingredient for the Lichnerowicz equation  we find
\begin{equation}
 W^2 = \wh \Lambda^2 V^2 = \wh \Lambda^ 2  K_{ij}K^{ij} = 18 \wh  \Lambda^ 2  J^2 \sin^2 \theta
\end{equation}
which agrees with \cite{BPS} except for the different $\Lambda$- scaling (which is already present in the respective  seed manifolds)
and that \cite{BPS} is restricted to the maximal case $\wt K = 0$.
We refer to that paper and to \cite{MK} for the discussion of the solutions.
\end{description}

\section{Marginally trapped surfaces} 
We finally locate and discuss  toroidal (marginally, outer)  trapped surfaces (MTSs, MOTSs)  as well as marginally trapped regions (MTRs) in our data. Before doing so in Sect.\,5.3,  we recall in Sect.\,5.1 key definitions, a result  on "(non)-visibility'' of MTRs 
due to Chru\'sciel, Galloway and Ling \cite{CGL} (reproduced below as Theorem 2)  which motivates our discussion,
 and in Sect.\,5.2   the situation in de Sitter spacetime and its perturbations.
 We adopt the notation of \cite{CGL} except that our tilded quantities refer to the physical spacetime, while compactifications
 are untilded. In particular, the spacetime is denoted by  $(\wt {\cal N}, \wt \ga_{\mu\nu})$ ($\mu,\nu = 0,1,2,3$) 
 and its compactification  by $({\cal N},  \ga_{\mu\nu})$.

\subsection{Definitions, results, motivation.}

\begin{defn}~
\begin{itemize}
\item  A {\bf marginally trapped surface (MTS)}  is a compact 2-surface for which   least one of the families of orthogonally emanating, future directed null  geodesics with tangents $\wt \l^+ $
and $\wt l^-$  has vanishing expansion ($\wt \Theta^+ =0$ or  $\wt \Theta^- = 0$).
\item
A  compact, connected spacelike hypersurface  is called a {\bf marginally trapped region (MTR)}  if its (only) boundary is a MTS with respect to the  outward normal $\wt l^+$.   In this case
the bounding MTS is called a {\bf marginally outer trapped surface (MOTS)}.
\end{itemize}
\end{defn}

\noindent
{\bf Remarks.}
\begin{enumerate}

\item
The above definition of a MOTS is in line with \cite{CGL} (which contains the more general definition of a weakly trapped region). However, it differs in general from others (e.g. Definition 2.1. of \cite{AMS}) as to the ``outer''  assignment.

\item A MOTS (and hence the boundary of a MTR) need not be connected.

\item A MOTS itself is {\it never} a marginally trapped region as the ``outer'' direction is ill-defined.

\item A MTR  need not contain any {\it outer trapped surfaces} defined by $\wt \Theta^{\pm} < 0$.
\end{enumerate}

The original motivation for studying such surfaces and regions comes from the singularity theorems. We recall in particular Hawking's classical theorem (Thm. 4, Sect. 8.2 of \cite{HE})  which asserts past geodesic incompleteness in spatially closed spacetimes that are at some stage future expanding and
 satisfy the strong energy condition.  However,  as the latter condition is violated in our $\Lambda$- vacuum case (while the dominant energy condition still holds for positive $\Lambda$), the conclusion need not hold; in fact de Sitter space  which is geodesically complete is an example. Nevertheless, as we shall see shortly, de Sitter space itself is awash with MTSs and MTRs. 
 We also recall that  Friedrich's stability results for de Sitter space \cite{HF2} indicate that, under ``weak'' energy conditions, a cosmological singularity theorem can  only hold under substantial modifications of the other requirements.

On the other hand,  Chru\'sciel, Galloway and Ling \cite{CGL} recently obtained results concerning  the  ``visibility" (from infinity) of   
MTSs and MTRs. The key differences to the singularity theorems are that only the null energy condition is required,  and
some  asymptotics compatible with de Sitter is assumed.  In precise terms, the result which concerns us here  reads as follows.\\

\noindent
{\bf Theorem 2. ((In)visibility of trapped regions from ${\cal J}$; slightly adapted Theorem 2.5  of \cite{CGL}.)}
{\it Consider a future asymptotically de Sitter spacetime $(\wt {\cal N}, \wt \gamma_{\mu\nu})$ which is  future causally simple and satisfies the null energy condition.  Then either the causal future of some set $\wt {\cal A} \subset  {\cal N}$ contains  all of infinity,
i.e. $J^{+}(\wt {\cal A},  {\cal N}) \supset {\cal J}^{+}$, or else there are no marginally  trapped regions  in
 $J^{+}(\wt {\cal A},  {\cal N}) \cap {\cal I}^- ({\cal J}^{+},  {\cal N})$.}\\

Turning to calculations,  we recall  the  decomposition of the expansion  on any 2-surface $\wt {\cal U}$ with mean curvature  $\wt H$, outer normal $\wt n^i$, and induced metric $\wt q_{ij}= \wt g_{ij} - \wt n_i \wt n_j$ in terms of the data, viz.

\begin{equation}
\label{expgen}
\wt \Theta^{\pm} = \pm \wt H + \wt Q =   \pm  \wt \nabla_i \, \wt n^i +   \wt q^{ij} \wt K_{ij} =   \pm  \frac{1}{\sqrt{\wt  g}}\, \partial_i \left( \sqrt{\wt g} \,\wt n^i \right)  +  \wt q^{ij} \wt K_{ij} \qquad   \mbox{on} ~  \wt {\cal U }.
\end{equation}

We now restrict ourselves to MOTSs of toroidal topology.  Such MOTSs have been found and studied before, in particular in asymptotically flat  $\Lambda = 0$-vacuum
data \cite{SH} as well as  in closed Friedmann-Lema\^{i}tre-Robertson-Walker spacetimes \cite{FHO, MX}.
  We remark that  topology results  (cf Lemma 9.2 of \cite{AMS}, or  \cite{RN}) 
   imply that MOTSs  which are {\it stable}  with respect to their outward normals within  their defining MTRs, 
(as defined e.g. in Definition 5.1. and Proposition 5.1. of  \cite{AMS}) must be  spherical.
Therefore, toroidal MOTSs must be {\it strictly unstable}  in the sense that the lowest eigenvalue of the stability operator 
(cf Definition 3.1 of \cite{AMS}) must be negative; this will be used in the discussion of Conjecture 3 in the next subsection.

\subsection{Toroidal MOTSs and MTRs in de Sitter spacetime} 

We next determine toroidal MOTSs on "standard'' CMC slices of de Sitter space, by which we mean 
$\sigma = \text{const}$, slices of  (\ref{deSi}). The induced metric and extrinsic curvature of such a slice read

\begin{equation}
\wt g_{ij} = \frac{3}{\Lambda \cos^2 \sigma}\, dS^2_{ij},  \qquad
\wt K_{ij} = \sqrt{\frac{3}{\Lambda}}\frac{ \sin \sigma}{ \cos^2 \sigma} dS^2_{ij},
\end{equation}

where $dS_{ij}^2$ is the standard metric on the unit three-sphere.
Hence  the mean curvature is  $\wt K =  \wt K_{ij} \wt g^ {ij} =   \sqrt{3 \Lambda}\sin \sigma$.

Restricting ourselves now to toroidal surfaces of the form $\tau= \tau_0 = \text{const}$ in the coordinates $(\ref{tau})$,
we obtain

\begin{eqnarray}
\label{MTS}
\left. \wt \Theta^{\pm} \right \vert_{\tau_0} & = &  \left. \left(\pm \wt H + \wt q^{ij} \wt K_{ij} \right) \right \vert_{\tau_{0}}   =
 \sqrt{\frac{\Lambda}{3}}\left[\pm \frac{  \cos \sigma}{\sin (2 \tau)} \frac{\partial}{\partial \tau}
  \left( \sin (2 \tau)  \right) +  2 \sin \sigma  \right]_{\tau_0}   =  \nonumber \\
& =&
2  \sqrt{\frac{\Lambda}{3}} \left( \pm \cos \sigma \cot (2 \tau_0)  +  \sin \sigma \right).
\end{eqnarray}
This implies that 

\begin{equation}
\label{deSMOTS}
\wt \Theta^{\pm} =0~~~ \mbox{iff}~~~ \tau = \tau_0^{\pm} = \pm \sigma/2 + \pi/4;
\end{equation}

 in particular  MOTSs exist for all times,  i.e. for all  $\sigma \in  (- \pi/2, \pi/2)$. The corresponding MTRs are given by
  $\tau \in [0, \tau_0^{+}]$ and   $\tau \in [ \tau_0^{-}, \pi/2]$, respectively;  curiously, neither region contains toroidal outer trapped surfaces.
  
As to applying Theorem  2 in this setup it should be kept in mind  that the set $\wt {\cal A}$ can in particular
be chosen to be a MTS, but alternatively to be a MTR, while the conclusion refers to  MTRs  in either case.

We first recall from \cite{CGL} the example of the time-symmetric case $\sigma_0 = 0$.  We take $\wt {\cal A}$ to be the Clifford torus at $\tau = \pi/4$ which satisfies $\wt \Theta^{\pm} = 0$. We find that  $J^{+}(\wt {\cal A}, {\cal N})$ contains ${\cal J}^{+}$ given by $\sigma = \pi/2$ in the compactification $ {\cal N}$; in fact it contains all slices $\sigma \in [\pi/4, \pi/2]$. These latter slices also contain MTRs, as  determined after Equ.  (\ref{deSMOTS}), while   $J^{+}(\wt {\cal A}, {\cal N}) \cap (\sigma = const)$  does not contain any MTRs for $\sigma \in [0, \pi/4)$. This is obviously consistent
with Theorem 2.  A similar behaviour is found for MOTSs $\wt {\cal A}$ given by (\ref{deSMOTS}) on any slice  $\sigma_0 \in (0, \pi/6]$: 
$J^{+}(\wt {\cal A}, {\cal N})$ contains all slices  $\sigma \in [3\sigma_0/2 + \pi/4, \pi/2]$, and only for such slices  
 $J^{+}(\wt {\cal A},  {\cal N}) \cap (\sigma = const)$  contains MRTs. On the other hand, for MOTSs on slices   $\sigma_0 \in ( \pi/6, \pi/2)$,  $J^{+}(\wt {\cal A},  {\cal N}) \cap (\sigma = const)$ contains neither   ${\cal J}^{+}$ nor any MTRs, again
 in agreement with Theorem 2.

Needless to say, one would like to have a more interesting example for this Theorem. A natural candidate would be a perturbation of 
de Sitter.  In fact Friedrich's stability result, Theorem 3.3  of \cite {HF2} together with  remark 3.4, asserts that, roughly speaking,  the compactification survives small perturbations of the data, which is a prerequisite in order for Theorem 2 to apply.  This motivates the following

\begin{conj} Under  perturbations of de Sitter data which preserve its global stucture  according to  Friedrich's  stability result Theorem 3.3. of \cite{HF2}, the toroidal  marginally outer trapped surfaces and marginally outer trapped regions remain close to those of de Sitter as determined above.  
\end{conj}

The difficulty of proving such a statement is that, as mentioned at the end of the previous subsection, toroidal MOTSs 
must be {\it strictly unstable} in the present $\Lambda-$ vacuum case. On the other hand,  {\it strict stability}   guarantees  the persistence of MOTSs under small perturbation of the data. This follows, via an implicit function argument, 
 from a slight adaption of Theorem 9.1 of  \cite{AMS}.  The same could be proven, by the same method,  in the present {\it strictly unstable} case provided the adjoint of the stability operator (Definition 3.1 of \cite{AMS}) had a  trivial kernel.  The latter, however, is unknown in the general setting as discussed above.
Below we will revisit  Conjecture 3 in the context of the special data constructed in Sect.\,4, without giving a proof either.   

\subsection{Toroidal MOTSs and MTRs in our data}

We now track toroidal MOTSs in the $\mathbb{U}(1) \times \mathbb{U}(1)$ -symmetric  data constructed in Sect.\,4  on $\mathbb{S}^3$.
We restrict ourselves to the maximal case $\wt K = 0$. As before the tori are given by $\tau = \tau_0 =  \text{const}$ in the coordinates (\ref{tau}), but now we include the  momentum  of Lemma  4.  For the last term in (\ref{expgen}), we obtain from  (\ref{BKT3}) and (\ref{sc})
\begin{equation}
\label{qk}
 \wt q^{ij} \wt L_{ij} =  \frac{2 }{3\, \phi^6} \left( u \|\Om\|^2 +  v \|\Th\|^2  \right) =
 \frac{2 }{3\, \psi^6} \left(\frac{\Lambda}{3}\right)^{3/2} \left( u \|\Om\|^2 +  v \|\Th\|^2  \right) =
2 \sqrt{ \frac{\Lambda}{3}}  \frac{c}{ \psi^6}
 \end{equation}
 which defines  $c$ {\it as a constant on $\mathbb{S}^3$}, in particular $c$ does not depend on the torus $\tau = \tau_{0}$.

Hence (\ref{expgen}) can be rewritten on $\wt {\cal U}$ in the coordinates (\ref{tau}) as
\begin{equation}
\label{exptor}
 \wt \Theta^{\pm}  =   \pm  \frac{1}{\phi^6 \sin (2\tau)} \,\frac{\partial}{\partial \tau} \left( \phi^4  \sin (2\tau)  \right) +
  \wt q^{ij} \wt L_{ij} =
  \pm  \frac{1}{\psi^6 \sin (2\tau)} \sqrt{ \frac{\Lambda}{3}} \,\frac{\partial}{\partial \tau} \left( \psi^4  \sin (2\tau)   \right)
 + 2 \sqrt{ \frac{\Lambda}{3}} \frac{ c}{\psi^6},
\end{equation}

and the condition for $\tau = \tau_0^{ \pm}$ to be a MOTS becomes

\begin{equation}
\label{MTS1}
\left[\frac{1}{\sin (2 \tau)} \frac{\partial}{\partial \tau}   \left(  \psi^4  \sin (2 \tau) \right) \right]_{\tau_0^{\pm}}    =  \mp 2 c.
\end{equation}

 We finally restrict ourselves to the special cases singled out in Remark 3 after  Theorem 1  and further
 elaborated in the previous section,  namely $\Lambda$-Taub-NUT ($v = 0$)   and the parallel ($u\, v = 1$) and orthogonal ($u\, v = -1$) cases. Furthermore  we adopt the choice (\ref{choice}), which gives $\|\Om\| = \sqrt{3/\Lambda}$ and $\|\Th\| = u \sqrt{3/\Lambda}$ .
 We find from the definition (\ref{qk}) that the respective constants $c$ take the values

 \begin{equation}
 \label{cs}
c_{\Join} = \frac{u}{3}   \qquad c_{\parallel} =  \frac{2 u}{3}   \qquad  c_{\bot} = 0.
 \end{equation}

In the following  closer  analysis of toroidal MOTSs  in the above cases we restrict ourselves to  the $\wt \Theta^+ = 0$ ones;
the case $\wt \Theta^- = 0$ involves some sign changes.

 \begin{description}
 \item[ $\Lambda$-Taub-NUT:]
Recall that here $W_{\Join}$ is constant given by the first of (\ref{3Ws}) and therefore $\psi_{\Join}$ is the
constant determined by (\ref{const}).
From  (\ref{MTS1}) and (\ref{cs})  the condition for a torus  $\tau = \tau_{\Join}$ to be  a MOTS then reads
\begin{equation}
\label{NUT}
\psi_{\Join}^4 =  - \frac{u}{3}  \tan (2 \tau_{\Join}) .
\end{equation}
Using (\ref{const}) and (\ref{3Ws}) we can  eliminate either $\psi_{\Join}$ or  $u$ to  obtain
\begin{equation}
\label{NUTMOTS}
\frac{u}{3} =    \cot(2 \tau_{\Join}) [4 \cot (2 \tau  _{\Join}) -1],  \qquad
\cot (2 \tau_{\Join}) = \pm \frac{ \sqrt{1 -\psi_{\Join}^4}}{2}.
\end{equation}
This calculation is  interpreted as follows. Recall from (\ref{const}) that, for any  $u$ with $u^2 \in (0, 1/3)$,
there are precisely two values for $\psi_{\Join} \in (0,1)$ which yield a stable and an unstable ``Premoselli pair''
of data. Either data have {\it precisely one} MOTS at $\tau_{\Join}$ given by the second equation in
(\ref{NUTMOTS}), where the sign has to be chosen such that  $u \tan (2 \tau _{\Join})  < 0$ by virtue of  (\ref{NUT}).
We now recover a behaviour analogous to the  de Sitter case:   (\ref{exptor}) implies that each torus given by
$\tau \in (0, \tau _{\Join})$ is {\it outer untrapped}  in the sense that $\wt \Theta^+ > 0$;    on the other hand, the region
covered by these tori is called a MTR  according to Definition 8.

\item[The parallel case:]  In the  previous section we determined numerically the stable and the unstable branches of solutions $\psi_{\parallel}$ of the  Lichnerowicz equation (\ref{lichx}) with $W_{\parallel}$ from (\ref{3Ws}).  Solving now also the MOTS  equation,
namely the  $\wt \Theta^+$ part of  (\ref{MTS})  with the choice (\ref{cs}) numerically reveals a behavior which is 
qualitatively the same as in the   previous $\Lambda$-Taub-NUT case: In particular we find {\it precisely one MOTS} $\tau =  \tau_{\parallel}$ on each branch.
We remark that for the unstable branch, the small-$u$ approximation \eqref{psi_pert} gives $\cos (2\tau_{\parallel}) = 2u/3 + O(u^2) $.

\item[The orthogonal case:]
The numerical solutions of the Lichnerowicz equation  (\ref{lichx}) now involve  $W_{\bot}$ from (\ref{3Ws}).
Since $c_{\bot}$ vanishes from  (\ref{cs}), the MOTS equation (\ref{MTS}) becomes
\begin{equation}
\label{MTSorth}
 \left.  4\,  \psi_{\bot}^{-1} \,  \frac{d \psi_{\bot}}{dx} \right\vert_{x_{\bot}}  =  \cot 2 \tau_{\bot}.
\end{equation}
A numerical analysis now shows that the respective sides of (\ref{MTSorth}) have different signs
unless  both vanish. Hence we are left with  the  Clifford  torus at $\tau_{\bot} = \pi/4
$ as only MOTS, like in  the time-symmetric de Sitter case described earlier.
\end{description}

To conclude, we found numerically  toroidal MOTSs in all $\mathbb{U}(1) \times \mathbb{U}(1)$-symmetric,
maximal $\Lambda$ -Taub-NUT,  ``parallel'' and ``orthogonal'' data. While in the first two cases there is a unique 
$\wt \Theta^{+} = 0 $ and $\wt \Theta^{-} = 0$- pair of  different MOTSs,  these MOTSs coincide at the Clifford torus in the latter case.
Being boundaries of MTRs, all these MOTSs qualify in principle as tests for the (non-)visibility theorem  of \cite{CGL}
quoted above as Theorem 2.    Clearly,  the constructed data are in general unlikely to satisfy the ``cosmic-no-hair''-type
requirement of this theorem, namely an evolution towards a causally simple asymptotically de Sitter spacetime 
$(\wt {\cal N}, \wt \gamma_{\mu\nu})$. We rather  return now to the perturbative setting of  Conjecture 3. 
For our special families of data this means that we need to restrict ourselves {\it both to small  $u$ as well as to the unstable solutions of the Lichnerowicz equation}, (in the sense of Definition 5 above), since the stable ones go to zero for $u \rightarrow 0$.
Clearly, $\Lambda-$ Taub NUT can for small NUT parameter  be interpreted   as a perturbation  of de Sitter, 
cf \cite{FB}. On the other hand, understanding  the structure of toroidal MOTSs and MTRs  in the other cases could 
be achieved by generalizing  the calculations of this subsection from the maximal to the CMC-case. 
We leave this to future work. \\ 

 \noindent
 {\large\bf Acknowledgements.}
We are grateful to Piotr Chru\'sciel, Dmitry Pelinovsky, and Bruno Premoselli  for helpful discussions and correspondence.
We also thank the referee for useful comments which led to improvements.  

The research of P.B. and W.S. was supported in part by the Polish National Science Centre  grant no. 2017/26/A/ST2/530.
W.S. also acknowledges support by the John Templeton Foundation Grant ``Conceptual Problems in Unification Theories'' (No. 60671).

\section{Appendix}

{\large\bf  $\Lambda$-Taub-NUT data}\\

The $\Lambda$-Taub-NUT metric can be written as  (cf. e.g. \cite{FB, OP})
\begin{equation}\label{Taub}
\wt \gamma_{\mu\nu} = \frac{3 D}{\Lambda}\left[ - \frac{1}{f(t)}t_{,\mu} t_{,\nu} + f(t)\Om^1_{\mu} \Om^1_{\nu} + (1 + t^2)(\Om^2_{\mu} \Om^2_{\nu} + \Om^3_{\mu} \Om^3_{\nu})
\right]\,,
\end{equation}
where the 1-forms $\Om^A_{\mu}$ are related to the vectors (\ref{om}) via  $\Om^A_{\mu}dx^{\mu} =  \, g_{ij}\,\Om^{A\,j} dx^i$,
and
\begin{equation}
f(t) = \frac{D t^4 + 2 (3 D - 2)t^2 + C t + 4 - 3D}{1 - t^2}
\end{equation}
 with constants $C$ and $D > 0$ so that $f > 0$.
We remark that the relation to the 1-forms $\omega^A_{i}$ of \cite{OP} is  $2 \, \Om^A_{\mu}dx^{\mu}  = \omega^A _{i} dx^i$
where the coordinates are related via $\tau = \theta/2$, $\ga = (\psi- \phi)/2$ and $\xi = (\psi +  \phi)/2$.

Note also that de Sitter spacetime is obtained for $C = 0,\,D = 1$ in the form
\begin{equation}
\label{deSi}
\wt \gamma_{\mu\nu} = \frac{3}{\Lambda \cos^2 \sigma}\,\left(- \sigma_{,\mu} \sigma_{,\nu} + \Om^1_\mu  \Om^1_\nu +
\Om^2_\mu \Om^2_\nu + \Om^3_\mu \Om^3_\nu\right)\,,
\end{equation}
where $t = \tan \sigma$.\newline
The intrinsic metric of $t = t_0$ is given by
\begin{equation}
\wt{g}_{ij}(t_0) = \frac{3 D}{\Lambda}[f(t_0) \Om^1_{i} \Om^1_{j} + (1 + t_0^2)( \Om^2_{i} \Om^2_{j} +
 \Om^3_{i} \Om^3_{j})]
\end{equation}
and the extrinsic curvature by
\begin{equation}
\wt{K}_{ij}(t_0) = \sqrt{\frac{3 D f(t_0)}{\Lambda}}\,\left[\frac{1}{2} f'(t_0) \Om^1_{i} \Om^1_{j} +
t_0( \Om^2_{i} \Om^2_{j} + \Om^3_{i} \Om^3_{j} )\right].
\end{equation}
The slice $t = t_0$ is conformal to the standard $\mathbb{S}^3$ iff $f(t_0) = 1 + t_0^2$. This leads to a relation  between $t_0$ and
the parameters $C$ and $D$ which we do not give explicitly.
For  the mean curvature of the spherical surfaces we obtain
\begin{equation}
\wt K = \wt K_{ij} \wt g^{ij} = \sqrt{\frac{\Lambda}{3D (1 + t_0^2)}} \left( \frac{1}{2} f' (t_0) + 2 t_0 \right).
\end{equation}
For a more detailed discussion we restrict ourselves to maximal slices (still with round metrics), which
satisfy $f'(t_0) = -4 \, t_ 0$. A computation shows that
\begin{equation}
C = 2 t_0\, \frac{t_0^4 + 6 t_0^2 - 3}{1 + t_0^2},\hspace{1cm}D = \frac{1 - t_0^2}{1 + t_0^2}.
\end{equation}
Assuming without loss that  $0 \leq t_0 < 1$,  the necessary and sufficient condition for the existence of such a $t_0$ is
\begin{equation}\label{C}
C = 4 \sqrt{\frac{1 - D}{1 + D}}\,\frac{1 - 2 D - 2 D^2}{1 + D}.
\end{equation}
and we are left with
\begin{equation}\label{family}
\wt{g}_{ij}(t_0) = \frac{6 D}{\Lambda (1 + D)} g_{ij},\hspace{1cm}\wt{K}_{ij}(t_0) =  3 \sqrt{\frac{3D}{2 \Lambda}}\,\frac{\sqrt{1 - D}}{1 + D}\,
\left(2 \Om^1_i \Om^1_j - \frac{2}{3} g_{ij}\right).
\end{equation}

One easily checks that this family of initial data  is a map from $0 < D \leq 1$ to solutions of the initial value constraints (\ref{con}), as it has to be, and this map is injective. To make contact with case I. of Theorem 1 in Sect. 3.  note that
\begin{equation} \label{ident}
\phi^2 = \sqrt{\frac{6 D}{\Lambda (1 + D)}},\hspace{1cm}u^2 =  \frac{9 D^2 (1 - D)}{(1 + D)^3}
\end{equation}

where we have used (\ref{BKT3}) with the choice $v=0$ and  (\ref{choice}).
From the second relation in (\ref{ident}) we see that each $u^2 \in [0,\frac{1}{3})$ has 2 inverse images $D \in [0,1]$, and this corresponds precisely to the (at least) 2 solutions of the Lichnerowicz equations predicted by Premoselli's theorem.

In any case we have shown that, with $u^2$ given as above for $0 < D \leq 1$,  our case I  of Theorem 1 
evolves into a $\Lambda$ - Taub - NUT metric with $C$ given by (\ref{C}).
We finally notice that, for $D$ close to $1$ (which implies $C$ close to $0$)  they must have regular future and past infinity as a consequence of Friedrich's stability result Theorem 3.3 of \cite{HF2}. As to the global structure of the general case we refer to
 \cite{FB}.


\begin{thebibliography}{00}

 \bibitem{AMS}
Andersson L, Mars M and Simon W 2008
Stability of marginally outer trapped surfaces and existence of marginally
outer trapped tubes
{\it Adv. Theor. Math. Phys.}~{\bf 12} 853


\bibitem{BK} Beig R and Krammer  W  2004
Bowen-York tensors {\it Class. Quantum Grav.}~{\bf 21} 73

\bibitem{FB}  Beyer F  2008 Investigations of solutions of Einstein's field equations close to
$\lambda$-Taub-NUT {\it Class. Quant. Grav.}~{\bf 25}, 235005

\bibitem{BPS} Bizo\'n P, Pletka S and Simon W  2015
Initial data for rotating cosmologies
  {\it Class. Quantum Grav.}~{\bf 32} 175015

\bibitem{BY} Bowen J M and York J 1980
Time-asymmetric initial data for black holes and black-hole collisions
{\it Phys. Rev. D}~{\bf 21} 2047

\bibitem{BL} Brezis H  and Li Y 2006
Some nonlinear elliptic equations have only constant solutions
{\it J. Partial Diff. Eqs.}~{ \bf 19} 208

\bibitem{PC} Chru\'sciel P T Cauchy problems for the Einstein equations: An Introduction, 
available under : {\it My lecture notes on the Cauchy problem}
{\tt https://homepage.univie.ac.at/piotr.chrusciel/teaching/Cauchy/Cauchy.html}
(unpublished)

\bibitem{CGL} Chru\'sciel P T,   Galloway  G L and Ling E 2018
Weakly trapped surfaces in asymptotically de Sitter spacetimes
{\it Class.  Quantum Grav.}~{\bf  35} 135001

\bibitem{CG} Chru\'sciel P T and Gicquaud R 2017  Bifurcating solutions of the
Lichnerowicz equation  {\it Ann. Henri Poincar\'e}~ {\bf 18}  643



\bibitem{LD} Dupaigne L 2011 Stable Solutions of Partial Differential Equations {\it Monographs and Surveys in Pure and Applied Mathematics}~{\bf 143}  (Boca Raton: Chapman \& Hall/CRC)

\bibitem{FHO} Flores J L,  Haesen S and Ortega M 2010
New examples of marginally trapped surfaces and tubes in warped spacetimes
{\it Class.  Quantum Grav}~{\bf 27}  145021


\bibitem{HF2}  Friedrich H 1986
On the existence of n-geodesically complete or future complete solutions of Einstein's field equations with smooth asymptotic structure
  {\it  Comm. Math. Phys.}~{\bf 107}, 587

\bibitem{HE} Hawking S W and Ellis G F R 2006  The Large Scale Structure of Space-Time
(Cambridge: Cambridge University Press)

\bibitem{SH} Husa S 1996  Initial data for general relativity containing a marginally outer trapped torus
{\it Phys. Rev. D}~{\bf 54} 7311


\bibitem{JI}  Isenberg J 2014 The initial value problem in General Relativity
{\it The Springer Handbook of Spacetime } Ed.  A. Ashtekar and V.
 Petkov  (New York: Springer)



\bibitem{JLX} Jin Q,  Li Y and  Xu H 2008
Symmetry and asymmetry: the method of moving spheres
{\it  Adv. Diff. Equ.}~{\bf 13} ~no.7-8~ 601

\bibitem{FK} Kottler F 1918  ~\"Uber die physikalischen Grundlagen der Einstein'schen Gravitationstheorie,
{\it Annalen der Physik}~{\bf 56}, 401


\bibitem{LP} Lee J M and Parker T H 1987
The Yamabe Problem {\it Bull. Am. Math. Soc. (New Ser.)}~{\bf 17} no.1~  37

\bibitem{MK}  Mach P and Knopik J 2018
Rotating Bowen-York initial data with a positive cosmological constant
{\it Class. Quantum Grav.}~{\bf  35} 145002

\bibitem{MX} Mach P and Xie N 2017  Toroidal marginally outer trapped surfaces in closed Friedmann-Lema\^{i}tre-Robertson-Walker spacetimes: Stability and isoperimetric inequalities
{\it Phys. Rev. D}~{\bf 96} 084050


\bibitem{RN} Newman R P A C  1987
Topology and stability of marginal 2-surfaces
{\it	Class Quantum Grav}~{\bf 4} 277


\bibitem{MO} Obata M  1972  The conjectures of conformal transformations of Riemannian manifolds {\it  J Diff Geom}~{\bf 6}  247

\bibitem{OP} Osuga K and  Page D N 2017
A new way to derive the Taub-NUT metric with positive cosmological constant
{\it J  Math Phys}~{\bf 58} 082501


\bibitem{BP}  Premoselli B 2015 Effective multiplicity for the Einstein-scalar
field Lichnerowicz equation {\it Calc. Var.}~{\bf 53} 29

\bibitem{RS}  Schoen R 1989 Variational theory for the Total Scalar Curvature
Functional for Riemannian metrics and related topics
{\it Topics in calculus of variations} ed M Giaquinta {\it Lecture Notes in
Math} {\bf 1365} (New York: Springer) p 120

\bibitem{Rick} Schoen R 1984 Conformal deformation of a Riemannian metric to constant scalar curvature
{\it J Diff Geom}~{\bf 20}  479

%


\end{thebibliography}
\end{document}